\documentclass[a4paper, amsfonts, amssymb, amsmath, reprint, showkeys, nofootinbib, twoside]{revtex4-2}
\pdfoutput=1
\usepackage[english]{babel}
\usepackage[utf8]{inputenc}
\usepackage[colorinlistoftodos, color=green!40, prependcaption]{todonotes}
\usepackage[pdftex, pdftitle={Article}, pdfauthor={Author}]{hyperref} 
\usepackage[T1]{fontenc} 
\usepackage{comment} 
\usepackage{todonotes}
\usepackage{amsmath}
\usepackage{amssymb}
\usepackage{bm}
\usepackage{natbib}
\usepackage{bbold}
\usepackage{epsfig}
\usepackage{hyperref}
\usepackage{graphicx}
\usepackage{csquotes}
\usepackage{mathtools}
\usepackage{slashed}
\usepackage{color,colordvi}
\usepackage{soul}
\usepackage{xcolor}
\usepackage{multirow}
\usepackage{tabularx,ragged2e,booktabs}
\usepackage{comment}
\usepackage{slashed}
\usepackage{float}
\usepackage{subcaption}
\usepackage{amsthm}
\usepackage{mathtools}
\usepackage{physics}
\usepackage{xcolor}
\usepackage{graphicx}
\usepackage[left=23mm,right=13mm,top=35mm,columnsep=15pt]{geometry} 
\usepackage{adjustbox}
\usepackage{placeins}
\usepackage[T1]{fontenc}
\usepackage{lipsum}
\usepackage{csquotes}
\usepackage[normalem]{ulem}
\renewcommand{\(}{\left(}
\renewcommand{\)}{\right)}
\renewcommand{\{}{\left\lbrace}
\renewcommand{\}}{\right\rbrace}

\newcommand{\ord}[1]{\mathcal{O}\({#1}\)}
\newcommand{\msbar}{\overline{\text{MS}}}
\newcommand{\as}{\alpha_\mathrm{s}}

\newcommand{\ordas}[1]{\mathcal{O}\left(\alpha_s^{#1}\right)}
\newcommand{\GeV}{\,\mathrm{GeV}}

\newcommand{\MeV}{\,\mathrm{MeV}}

\newcommand{\hs}{\hspace{.4mm}}

\newcommand{\bs}{\hspace{1cm}}

\newcommand{\lqcd}{\Lambda_{\text{QCD}}}

\begin{document}
\title{Renormalization group improved determination of \texorpdfstring{$\as$}{}, \texorpdfstring{$m_c$}{}, and \texorpdfstring{$m_b$}{} from the low energy moments of heavy quark current correlators}

\author{M. S. A. Alam Khan}
    \email[Correspondence email address: ]{alam.khan1909@gmail.com}
    \affiliation{Centre for High Energy Physics, Indian Institute of Science, Bangalore 560 012, India}

\date{\today} 

\begin{abstract}
We determine $\as$, $m_c$, and $m_b$ using the relativistic quarkonium sum rule and the renormalization group summed perturbation theory (RGSPT). Theoretical uncertainties, especially originating from the variation of the renormalization scale, are considerably reduced for the higher moments. Our determinations using RGSPT are also found to be stable with respect to the use of $\msbar$ quark mass for the condensate terms. We obtain $\alpha_s^{\left(n_f=5\right)}(M_z)=0.1171(7)$, $\overline{m}_c=1281.1(3.8)\hs\MeV$, and $\overline{m}_b=4174.3(9.5)\hs\MeV$. 
\end{abstract}


\maketitle

\section{Introduction}
 The strong interaction in the standard model (SM) of particle physics describes the interactions of the quarks and gluon. These interactions are very precisely studied under quantum chromodynamics (QCD) which is non-perturbative at low-energy regions and has asymptotic freedom~\cite{Gross:1973id,Politzer:1973fx} at high energies. The QCD scale, $\lqcd$, is a scale parameter that separates these energy regimes. At low energies, where momentum transfer ($q$) is of the order of $\lqcd$, chiral perturbation theory (ChPT)  and lattice QCD are powerful methods to describe strong interactions. ChPT describes the interactions of pion and kaons while the lattice QCD computations are improving over the years, and now predictions even for the bottom quark systems are also available~\cite{FlavourLatticeAveragingGroupFLAG:2021npn}. The perturbative nature of QCD at high energies ($q\gg\lqcd$) allows one to use methods like operator product expansion (OPE) to systematically calculate the various quantities as an expansion of strong coupling constant ($\as$) by evaluating the Feynman diagrams appearing at different orders of $\as$. The OPE also parametrizes the non-perturbative physics in the condensates involving the quarks and gluon fields. These condensates can be calculated using lattice QCD and ChPT~\cite{Ioffe:2005ym}, Optimized perturbation theory (OPT)~\cite{Kneur:2020bph} or using powerful tools such as QCD sum rules~\cite{Shifman:1978bx,Shifman:1978by}. For more details about the applications of the QCD sum rules, we refer to Refs.~\cite{Dominguez:2018zzi,Narison:2022paf}. \par
The effective field theories (EFT) of the strong interactions play a key role in studying systems ranging from a few $\MeV$ to several $\GeV$. For reviews, we refer to Refs.~\cite{Ananthanarayan:2023gzw,Meng:2022ozq,Casalbuoni:1996pg,Brambilla:1999xf,Brambilla:2010cs,Brambilla:2004jw,Mannel:2020ups}. Since EFTs are formulated for a very specific energy range, they are sensitive to fewer parameters than full QCD. These features allow an efficient determination of the parameters of the SM using QCD sum rules with the experimental information taken as inputs.\par 
The low-energy moments ($\mathcal{M}_n^{X}$) of the current correlators, defined in Eq.~\eqref{eq:Moments_V}, are important quantities that can be theoretically calculated. The corresponding quantity is obtained from the experimental data on the resonances, which are only available for the vector channel (V). The moments for the pseudoscalar channel (P) can not be obtained from real experiments but can be obtained using the lattice QCD simulations~\cite{HPQCD:2008kxl,McNeile:2010ji,Maezawa:2016vgv,Nakayama:2016atf,Petreczky:2019ozv,Petreczky:2020tky}. From these simulations, the dimensionless quantities such as $\mathcal{M}_0^{P}$ and the ratios of the higher moments ($\mathcal{R}_n^{P}$), defined in Eq.~\eqref{eq:Def_R}, can be reliably obtained. These results are used in the determination of $\as$, bottom quark mass ($m_b$), charm quark mass ($m_c$), and the non-perturbative quantities such as the gluon condensates~\cite{Kuhn:2001dm,Ahmady:2004er,Chetyrkin:2009fv,Narison:2010cg,Chetyrkin:2010ic,Narison:2011xe,Dehnadi:2011gc,Beneke:2014pta,Dehnadi:2015fra,Erler:2016atg,Boito:2020lyp,Boito:2019pqp,Erler:2022mzd}. Other QCD sum rules-based determinations can be found in Refs.~\cite{Peset:2018ria,Signer:2007dw,Signer:2008da,Bodenstein:2011ma,Bodenstein:2011fv,Bodenstein:2010qx,Dominguez:2014pga,Kiyo:2015ufa,Narison:2019tym} and for recent lattice QCD determinations, we refer to Refs.~\cite{FermilabLattice:2018est,Komijani:2020kst}. \par
The determination of these parameters using traditional fixed-order perturbation theory (FOPT) series from the lower moments is dominated by experimental uncertainties, but higher moments are dominated by theoretical uncertainties. Theoretical uncertainties arise when parameters such as $\as$, quark masses ($m_q$), the gluon condensate ($\langle\frac{\as}{\pi} G^2\rangle$) are taken as input, and the renormalization scale ($\mu$) is varied in a certain range. Higher moments are more sensitive to the renormalization scale dependence and therefore dominated by its uncertainties. Also, the $\msbar$ definition of the quark mass for the vector channel, when used in the non-perturbative gluon condensate terms, gives unreliable determinations for the strong coupling and quark masses. This problem is cured by using the on-shell mass taken as input~\cite{Chetyrkin:2017lif,Dehnadi:2011gc}.\par 
In this article, we have addressed these issues by summing the running logarithm using the RGSPT. In this scheme, the running logarithm arising from a given order is summed to all orders in closed form using the renormalization group equation (RGE). This scheme has already been found to be useful in other processes in Refs.~\cite{Abbas:2012py,Ananthanarayan:2016kll,Ananthanarayan:2022ufx,Ahmady:2002fd,Ahmady:2002pa,Ananthanarayan:2020umo,Ahmed:2015sna,Abbas:2022wnz,Chishtie:2018ipg,Abbas:2022wnz,Abbas2}.\par It should be noted that there is already an existing $m_b$ determination using RGSPT by Ahmady et al. in Ref.~\cite{Ahmady:2004er}. Special emphasis was given to the scale reduction in the $\msbar$ and its conversion to the pole mass scheme and the $1S$ scheme. Since then, there has been a significant reduction in the uncertainties in the experimental moments and the value of the strong coupling constant. Also, the first four moments to four-loop ($\as^3$), and the quark mass relations to four-loop ($\as^4$) are now available. This information can be used to further reduce the theoretical uncertainties. With these advantages in hand, we take one step further and extend its application in the $m_c$ and $\as$ determinations. \par 
In section \eqref{sec:formulas}, we briefly discuss various quantities relevant to this article. In section \eqref{sec:RGmom}, we discuss the renormalization group (RG) improvement of the moments using RGSPT. Since the pseudoscalar and vector channel moments are available for the charm case, we use these moments in the determination of $m_c$ and $\as$ in sections \eqref{sec:mc_det} and \eqref{sec:as_det}, respectively. In section \eqref{sec:mb_det}, $m_b$ is obtained only from the vector moments. In section \eqref{sec:summary}, we provide our final determination, and the importance of the RGSPT is discussed in detail. The supplementary material needed in this article is presented in the appendix~\eqref{app:pert_coef} and \eqref{app:RGcoefs}.\par 
Before moving to the next section, it should be noted that we use the following numerical inputs from PDG~\cite{ParticleDataGroup:2022pth} are used in this article:
\begin{equation}
\begin{aligned}
    M_c&=1.67\pm0.07\GeV\,,\\
    M_b&=4.78\pm0.06\GeV\,,\\
    \as^{\left(n_f=5\right)}(M_Z)&=0.1179\pm0.0009\,,\\
    m_c(m_c)&=1.27\pm0.02\GeV\,,\\
    m_b(m_b)&=4.18\pm0.04\GeV\,.
\end{aligned}
\label{eq:num_input}
\end{equation}
and decoupling and the running of $\as$ is performed at the $\msbar$ scheme values of the charm and bottom quark masses using REvolver~\cite{Hoang:2021fhn} and RunDec~\cite{Herren:2017osy} packages.

\section{Theoretical inputs}\label{sec:formulas}

The normalized total hadronic cross-section ($R_{\overline{q}q}$), defined as:
\begin{align}
	R_{\overline{q}q}\equiv\frac{3s}{4\pi\alpha^2}\sigma\left(e^+e^-\rightarrow q\overline{q}+X\right)\simeq\frac{\sigma\left(e^+e^-\rightarrow q\overline{q}+X\right)}{\sigma\left(e^+e^-\rightarrow \mu^+\mu^-\right)}\,,
 \label{eq:Rratio}
\end{align}
is one of the most important observable sensitive to the quark mass ($m_q$). The inverse moment for the vector channel ($\mathcal{M}_q^{V,n}$), are derived from $R_{\overline{q}q}$ as:
\begin{align}
	\mathcal{M}_q^{V,n}=\int\frac{ds}{s^{n+1}}R_{q\overline{q}}\,.
 \label{eq:Moments_V}
\end{align}

It is evident from Eq.~\eqref{eq:Moments_V} that for higher moments, significant contributions come from low energy resonances. To quantify these contributions, theoretical inputs from the non-relativistic QCD (NRQCD)~\cite{Caswell:1985ui,Bodwin:1994jh} play a crucial role. Their results can be taken as input in the determination of the $m_c$ and $m_b$ Ref.~\cite{Signer:2007dw,Signer:2008da} and the sum rules are usually referred to as the non-relativistic sum rule. \par

 Using analyticity and unitarity, the moments are related to the coefficients of the Taylor expansion for the quark-heavy correlator evaluated around $s=0$ as:
\begin{align}
	\mathcal{M}_{n}^{V,\text{th}}=\frac{12\pi^2Q^2_q}{n!}\frac{d^n}{ds^n}\Pi^V(s)\Big|_{s=0}
\end{align}
where $Q_q$ is the electric charge, $s=\sqrt{q^2}$ is the $e^+e^-$ center of mass energy, and $\Pi_V(s)$ are the current correlators of two vector currents given by:
\begin{align}
	\left(s\hs g_{\mu\nu}-q_\mu q_\nu\right)\Pi^V(s)=-i \int dx e^{i q x}\langle0\vert T\{j_\mu(x)j_\nu(0)\}\vert0 \rangle\,,\nonumber
\end{align}
where, \begin{align}
	j_\mu=\overline{q}(x)\gamma^\mu q(x)\,.\nonumber
\end{align}
\par 
For the pseudoscalar channel, slightly different definitions are used in Ref.~\cite{Dehnadi:2015fra} and which is also adopted in this article. The pseudoscalar current correlator is defined as
\begin{align}
    \Pi^{P}(s)&\equiv i\int d\hs x e^{i\hs q\hs x}\langle 0\vert T\lbrace j_P(x)\hs j_P(0)\rbrace\vert 0\rangle\,,
    \label{eq:vacP}
\end{align}
where \begin{align}
    j_P=2\hs i\hs m_q\hs \overline{q}(x)\gamma^5 q(x)\,,
\end{align}
and the double subtracted polarization function is obtained from Eq.~\eqref{eq:vacP} as:
\begin{align}
    P(s)=\frac{1}{s^2}\left(\Pi^{P}(s)-\Pi^{P}(0)-s\hs\left[ \frac{d}{d\hs s}\Pi^{P} (s)\right]_{s=0}\right)\,,
\end{align}
from which the moments are obtained as:
\begin{align}
    \mathcal{M}^{P,\text{th}}_n(s)=\frac{12\pi^2 Q_q^2}{n!} \frac{d^n}{d\hs s^n}P(s)\Big|_{s=0}\,.
\end{align}
Theoretical moments are calculated using the OPE and have contributions from purely perturbative ($\mathcal{M}_n^{X,\text{pert}}$) as well as non-perturbative ($\mathcal{M}_n^{X,\text{n.p}}$) origin. Therefore, we can write the theoretical moments as follows:
 \begin{align}
\mathcal{M}_{n}^{X,\text{th}}=\mathcal{M}_n^{X,\text{pert.}}+\mathcal{M}_n^{X,\text{n.p.}}\,.
\label{eq:Def_MX}
 \end{align}
 The fixed order perturbative series for $\mathcal{M}_n^{X,\text{pert.}}$ have the following form:
\begin{align}
    \mathcal{M}_n^{X,\text{pert}}=m_q^{-2n}\sum_{i=0}T_{i,j}^X x^i L^j
    \label{eq:mom_fopt}
\end{align}
where $m_q\equiv m_q(\mu)$, $x\equiv\as(\mu)/\pi$ and $L\equiv \log(\mu^2/q^2)$. \par The $T_{i,0}^{X,Y}$ are RG inaccessible terms calculated using the perturbation theory by evaluating the Feynman diagrams appearing in a given order. Their numerical values are presented in appendix~\eqref{app:pert_coef}. Other $T_{i,j}^{X,Y}$ coefficients can be obtained using the RGE and are known as RG-accessible terms. 
The two-loop correction to $\mathcal{M}_n^{X,\text{pert}}$ are calculated in Ref.~\cite{Kallen:1955fb}, three-loops in Refs.~\cite{Chetyrkin:1995ii,Chetyrkin:1996cf,Boughezal:2006uu,Czakon:2007qi,Maier:2007yn}, the first four moments at four-loop (or $\as^{3}$)  from Refs.~\cite{Hoang:2008qy,Maier:2009fz}. Predictions for higher moments using the analytic reconstruction method can be found in Refs.~\cite{Greynat:2010kx,Greynat:2011zp} and are used in Ref.~\cite{Greynat:2012bxq} in the $m_c$ determination. Other predictions using Pad\'e approximants can be found in Ref.~\cite{Kiyo:2009gb}. A large-$\beta_0$ renormalon-based analysis for the low energy moments of the current correlators can be found in Ref.~\cite{Boito:2021wbj}.\par
 The $\mathcal{M}_n^{X,\text{n.p}}$ include the contributions from the condensate terms and has the following form:
\begin{align}
\mathcal{M}_n^{X,\text{n.p.}}=&\frac{1}{\left(2\hs m_q\right)^{4n+4}}\Big\langle\frac{\as}{\pi}G^2\Big\rangle_{\text{RGI}}\nonumber\\&\times\hs\left(T^{X,\text{n.p.}}_{0,0}+x(m_q) T^{X,\text{n.p.}}_{1,0}\right)+\ord{x^2}\,.
    \label{eq:def_cond}
\end{align}
 where, $T^{X,\text{n.p.}}_{i,0}$ are the perturbative correction as prefactors to the gluon condensate and are known to NLO~\cite{Broadhurst:1994qj} and can be found in the appendix~\eqref{app:pert_coef}. For the RG invariant  gluon condensate, we use the following numerical value~\cite{Ioffe:2005ym}:
 \begin{align}
     \langle\frac{\as}{\pi}G^2\Big\rangle_{\text{RGI}}=0.006\pm0.012\GeV^4\,.
 \end{align}
In addition, we also need quark mass relations to convert it from the $\msbar$ scheme to the on-shell scheme. These relations are now known to four-loops~\cite{Tarrach:1980up,Gray:1990yh,Fleischer:1998dw,Chetyrkin:1999qi,Marquard:2015qpa,Marquard:2016dcn}. The one-loop relation relevant for this article is given by:
\begin{align}
    m_q(\mu)=M_q \left(1-x(\mu)\left(\frac{4}{3}+\log\left(\frac{\mu^2}{M_q^2}\right)\right)\right)+\ord{x^2}\,,
\end{align}
which will be used in Eq.~\eqref{eq:def_cond} for the quark condensate terms.\par
From theoretical moments, defined in Eq.~\eqref{eq:Def_MX}, the ratio of the moments ($\mathcal{R}^{X}_n$) can be obtained as:
\begin{align}
\mathcal{R}^{X}_n\equiv\frac{\left(\mathcal{M}^{X}_n\right)^{\frac{1}{n}}}{\left(\mathcal{M}^{X}_{n+1}\right)^{\frac{1}{n+1}}}\,,
\label{eq:Def_R}
\end{align}
which are more sensitive to the $\as$ and less sensitive to the quark masses. The mass dependence arises only from the running logarithms present in the perturbative expansion. This quantity is very useful in the determination of the $\as$.\par With an introduction to these RG invariant quantities, we are in a position to discuss their RG improvement using RGSPT in the next section.

\section{RG Improvement of Moments} 
\label{sec:RGmom}
The FOPT expression for the $\mathcal{M}_n^{X,\text{pert.}}$, in Eq.~\eqref{eq:mom_fopt}, is a RG invariant perturbative expansion quark mass and $\as$. The evolution of the quark masses and $\as$ is dictated by their RGE. There are also some studies where $m_q(\mu)$ is expanded in the $\msbar$ scheme, and an extra scale ($\mu_m$) is introduced whose effects appear in $\ordas{2}$ in running logarithm. Although this procedure is very general, independent scale variations of $\left(\mu,\mu_m\right)$ give more renormalization scale uncertainty as the RG invariance of the moments $\mathcal{M}_n^{X,\text{pert.}}$ is broken in the case of the finite order results. Since this article only focuses on the RG improvement, we restrict ourselves to the single renormalization scale ($\mu$), known as a correlated choice of scale approach.\par  
    To obtain the closed for summed expression, we rewrite the perturbative series in Eq.~\eqref{eq:mom_fopt} as follows:
    \begin{align}
        \mathcal{M}_n^{X,\Sigma}= m_q^{-2\hs n}\sum_{i=0}x^i\hspace{.4mm}S_{i}(x \hspace{.4 mm} L)\,,
        \label{eq:ser_summed}
    \end{align}
    where the $S_{i}\left(x L\right)$ are the RG summed coefficients given by:
    \begin{align}
        S_{i}\left(x\hspace{.4mm}L\right)=\sum_{n=i}^{\infty} T^X_{n,n-i}  (x\hspace{.4mm}L)^{n-i}\,.
        \label{eq:summed_coefs}
    \end{align}
    Since, $\mathcal{M}_n^{X,\text{pert.}}$ is an observable, Eq.~\eqref{eq:mom_fopt} has a homogeneous RGE given by:
    \begin{align}
        &\mu^2 \frac{d}{d\mu^2}\mathcal{M}_n^X=0\,,\\
        \implies& \left(\beta(x) \partial_x+ \gamma_m(x) \partial_m+\partial_L\right)\mathcal{M}_n^X=0\,,
    \end{align}
    where $\beta(x)$ and $\gamma_m$ are the QCD beta function~\cite{vanRitbergen:1997va,Gross:1973id,Caswell:1974gg, Jones:1974mm,Tarasov:1980au,Larin:1993tp,Czakon:2004bu,Luthe:2016ima,Baikov:2016tgj,Herzog:2017ohr} and quark mass anomalous dimension~\cite{Tarrach:1980up,Tarasov:1982plg,Larin:1993tq,Vermaseren:1997fq,Chetyrkin:1997dh,Baikov:2014qja,Luthe:2016xec,Baikov:2017ujl} given by:
     \begin{align}
      \beta(x)&\equiv \mu^2\frac{d}{d\mu^2}x(\mu)=-\sum_i \beta_i x^{i+2} \,, \label{eq:beta_function}\\
 \gamma_m&\equiv \mu^2\frac{d}{d\mu^2}m_q(\mu)=-m_q(\mu)\sum_i \gamma_i\hs x^{i+1} \,. \label{eq:mass_anom}
    \end{align}
     Now, we can follow the steps described in Ref.~\cite{Ahmady:2002fd} by collecting coefficients corresponding to summed coefficients defined in eq.~\eqref{eq:summed_coefs}. This process results in a set of coupled differential equations for $S_{i}( x\hs L)$, which can be written in a compact form as:
    \begin{align}
        \sum _{i=0}^k \bigg[\beta _i &(\delta_{i,0}+w-1)  S_{k-i}'(w)\nonumber\\&+S_{k-i}(w) \left(-2 n \gamma_i+\beta _i (-i+k)\right)\bigg]=0\,
        \label{eq:summed_de}
    \end{align}
   where, $w\equiv 1-\beta_0\hs x\hs L$. The solutions for the above differential equation are presented in the appendix~\eqref{app:RGcoefs}. We can obtain various $\mathcal{M}_n^{X,\Sigma}$ from these solutions. It should be noted that the corresponding expression in the on-shell scheme is obtained by setting quark mass anomalous dimension $\gamma_i=0$.\par
\begin{figure}[H]	
\centering
		\includegraphics[width=.40\textwidth]{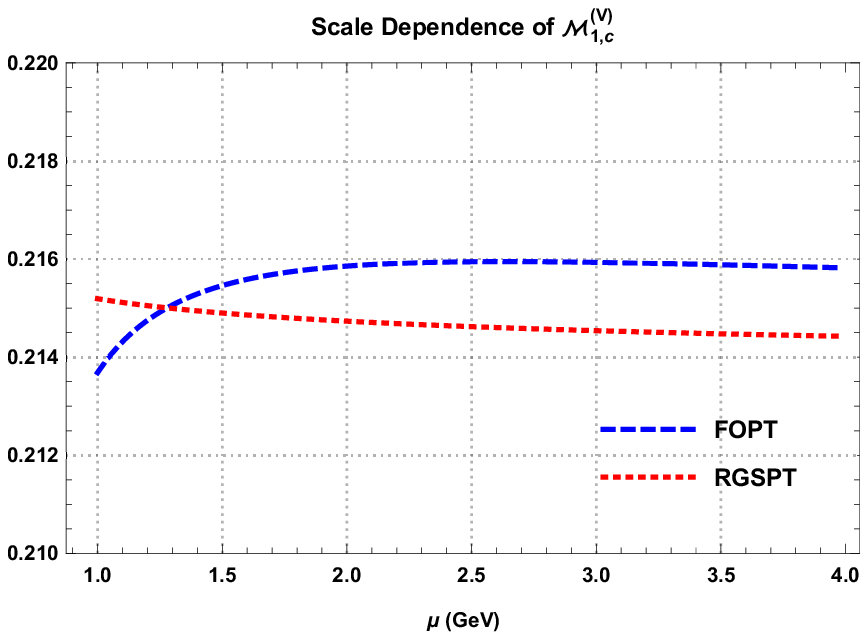}
		\includegraphics[width=.40\textwidth]{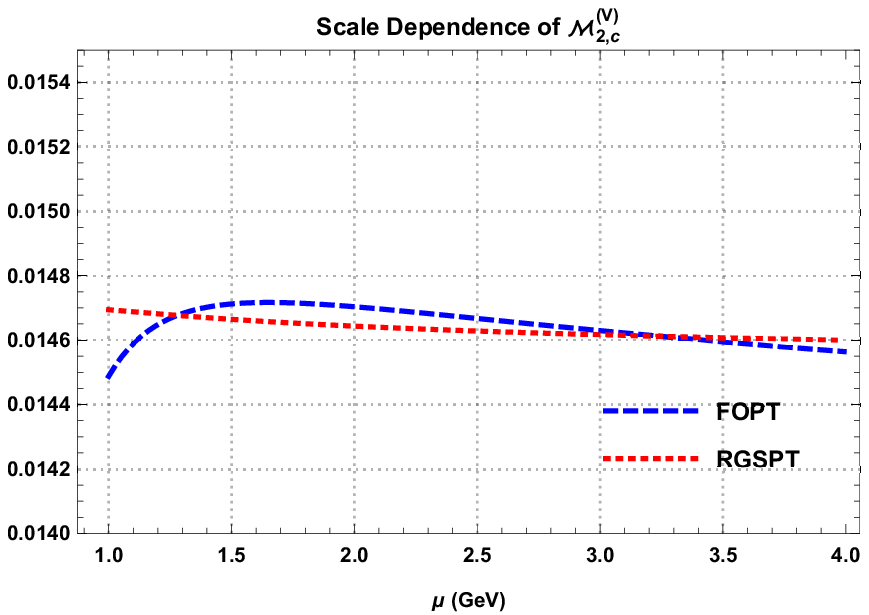}
	\includegraphics[width=.40\textwidth]{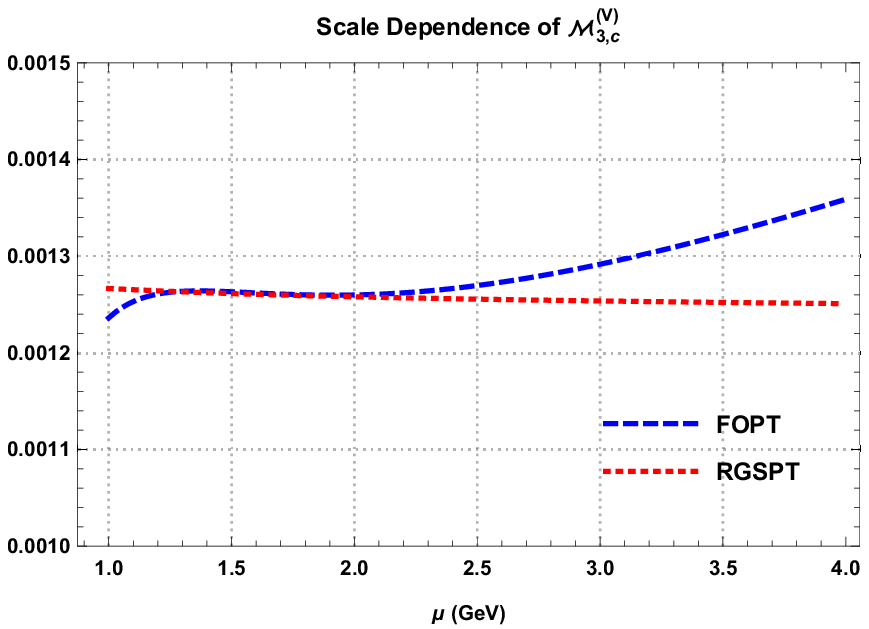}
	\includegraphics[width=.40\textwidth]{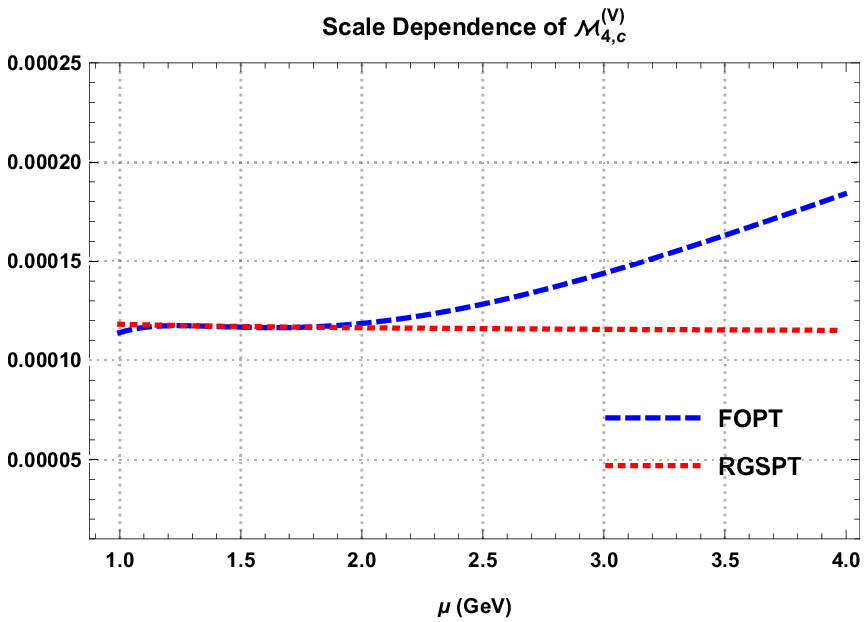}
\caption{\label{fig:MomV_c} Renormalization scale dependence of the first four vector moments for the charm quark.}
\end{figure}
\begin{figure}[H]\centering
		\includegraphics[width=.40\textwidth]{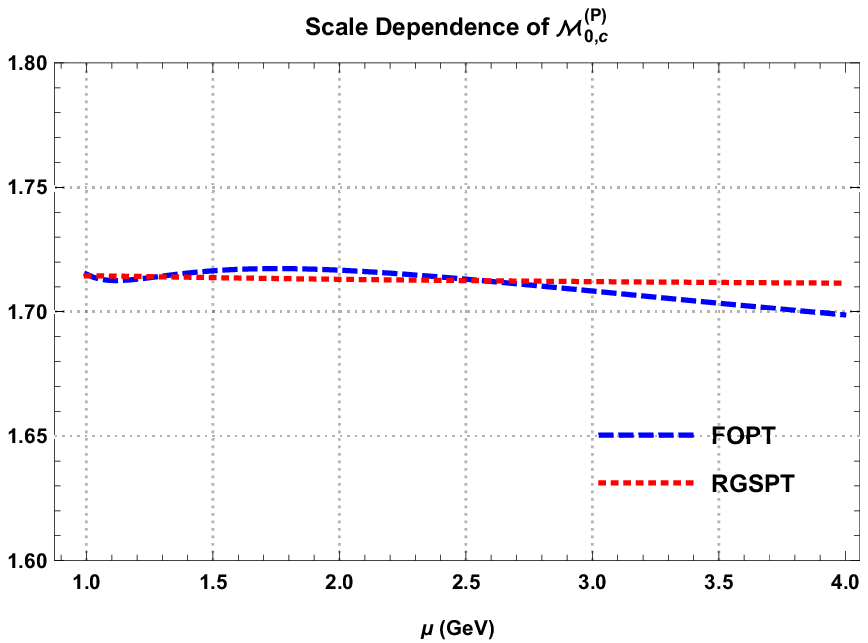}
		\includegraphics[width=.40\textwidth]{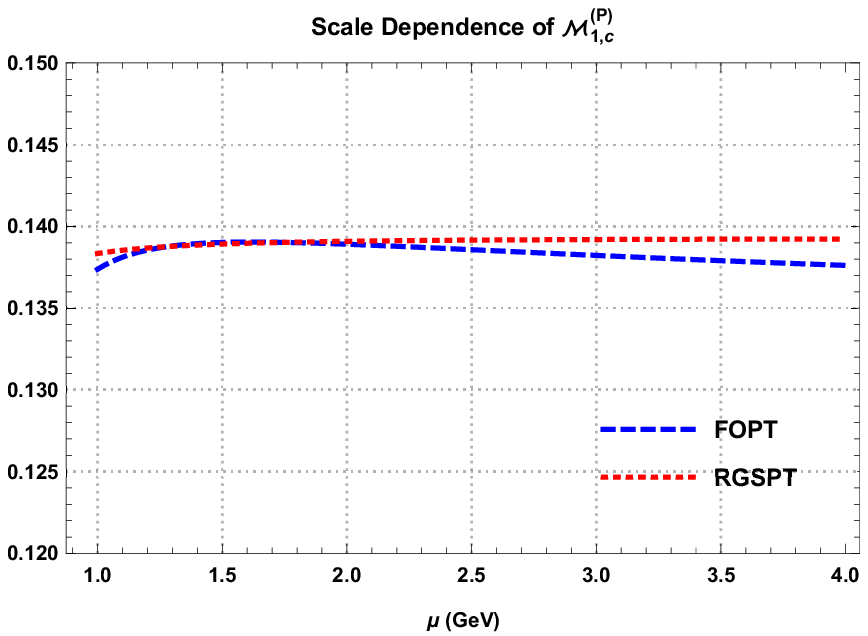}
	\includegraphics[width=.40\textwidth]{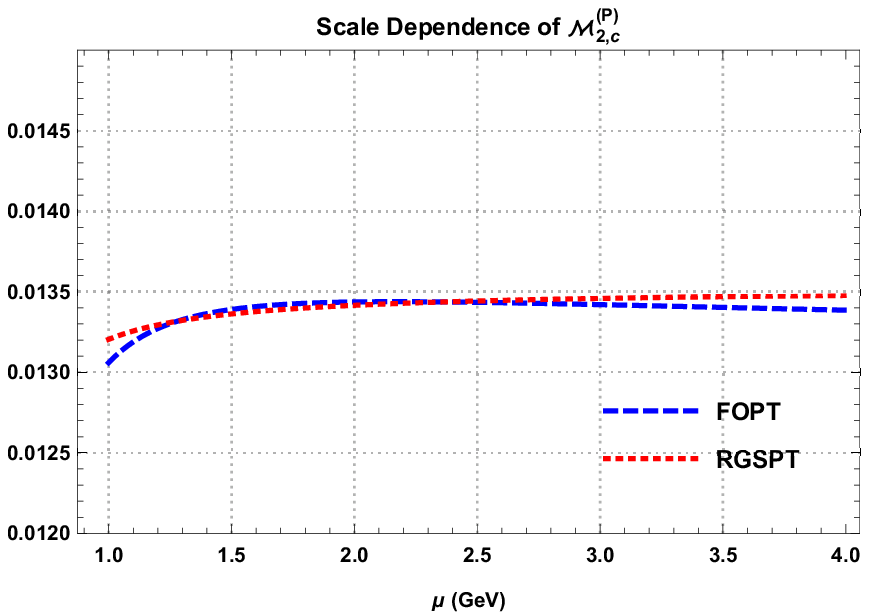}
	\includegraphics[width=.40 \textwidth]{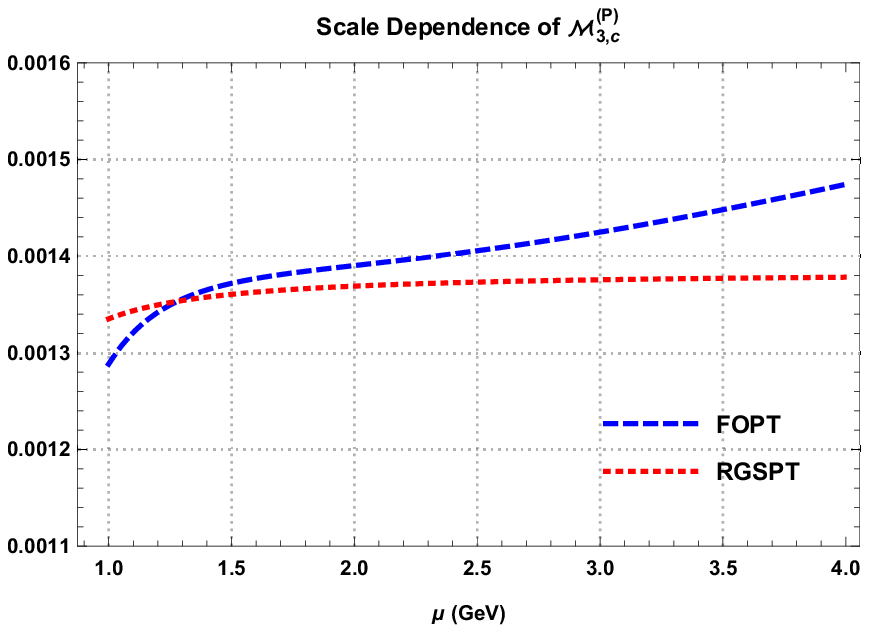}
\caption{\label{fig:MomP_c} Renormalization scale dependence of the first four pseudoscalar moments for the charm quark.}
\end{figure}
\begin{figure}[H]
\centering
		\includegraphics[width=.40\textwidth]{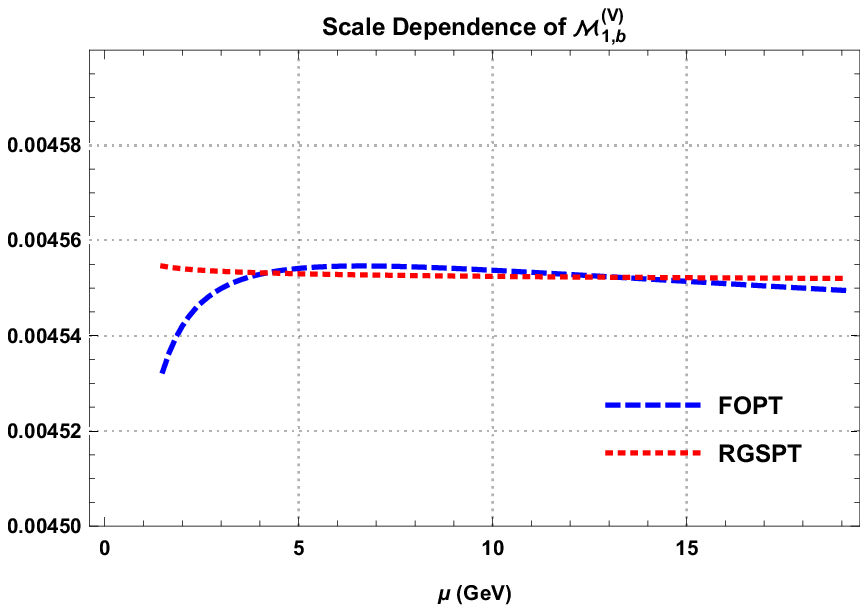}
		\includegraphics[width=.40\textwidth]{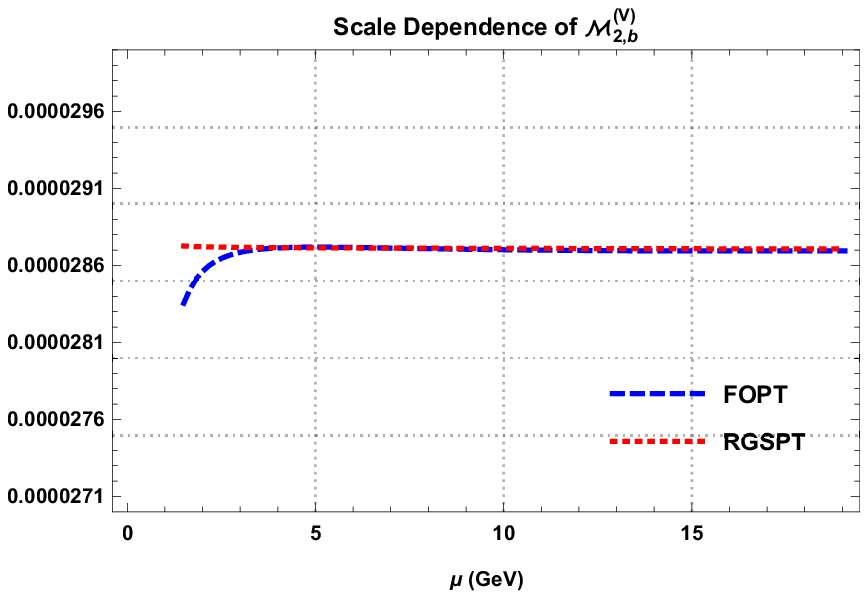}
	\includegraphics[width=.40\textwidth]{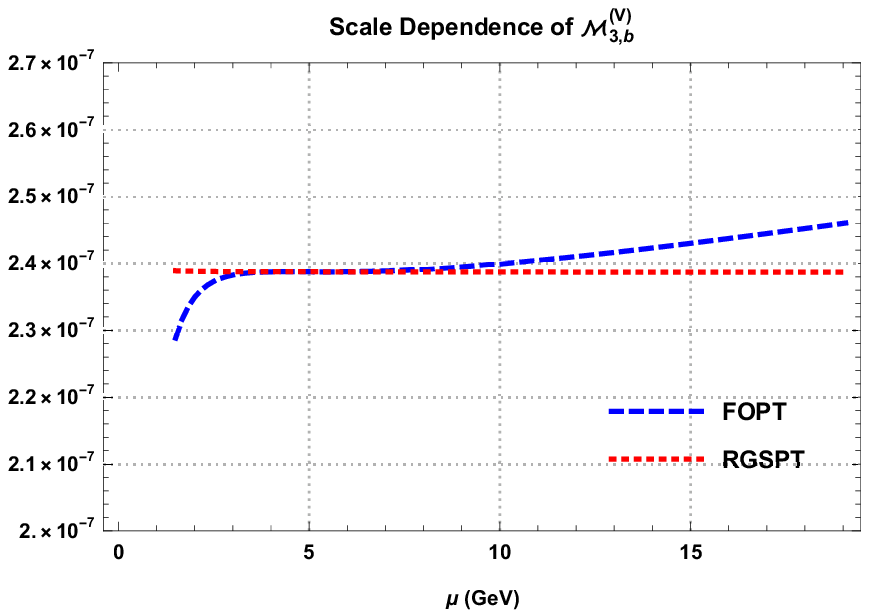}
	\includegraphics[width=.40\textwidth]{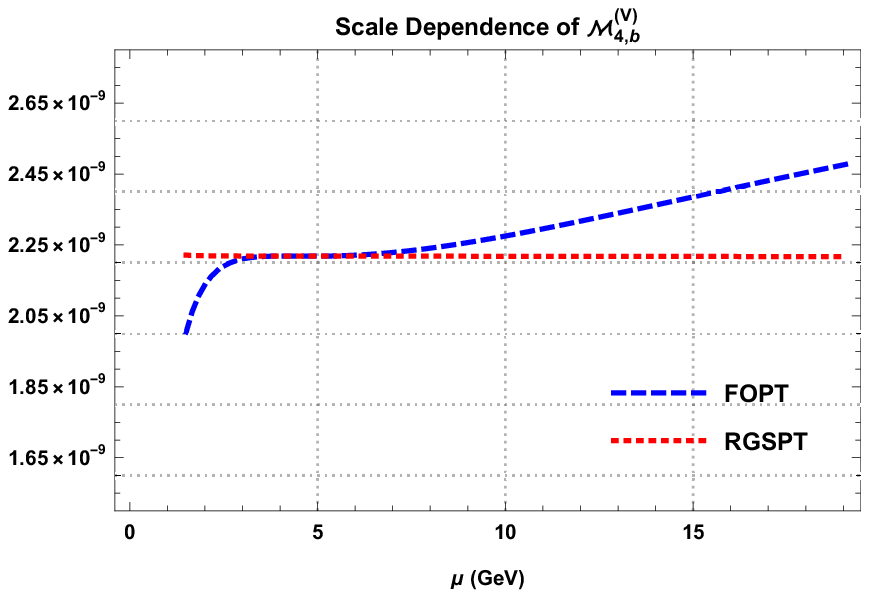}
\caption{\label{fig:Mom_b} Renormalization scale dependence of the first four vector moments for the bottom quark.}
\end{figure}

After RG improved perturbative series is obtained for different $\mathcal{M}^{X}_n$, we can study their scale dependence. For the charm moments, we take $\as^{\left(n_f=4\right)}(3\GeV)=0.2230$ and $m_c(3\GeV)=993.9\MeV$. For the bottom moments, we take $\as^{\left(n_f=5\right)}(10\GeV)=0.1780$ and $m_b(10\GeV)=3619.4\MeV$. These values are obtained from Eq.~\eqref{eq:num_input} using the REvolver package. The scale dependence of the first four moments for the vector and pseudoscalar channel for the charm case can be found in Fig.~\eqref{fig:MomV_c} and Fig.~\eqref{fig:MomP_c}, respectively. We only used vector moments for the bottom quark case, and the scale dependence can be found in Fig.~\eqref{fig:Mom_b}. It should be noted that the agreement of various moments in FOPT and RGSPT prescription occurs at the $\msbar$ value of the quark masses, i.e. $\mu=m_q(\mu)=m_q(m_q)$. At this particular scale, the RGSPT expressions reduce to FOPT expressions. It is evident from these figures that the RGSPT has better control of the scale variations compared to the FOPT. For the vector moments, the third and fourth moments in the FOPT scheme are very sensitive to scale variations and contribute to a large theoretical uncertainty even though their experimental values are known more precisely. With these advantages in hand, we have used FOPT and RGSPT in the determinations of the $\as$, $m_c$, and $m_b$ in the next sections. 
\section{Charm mass determination}
Charm quark is very interesting for the low as well high energy regime of the QCD. It is not heavy enough that heavy quark effective theory can be used for precise prediction nor close enough $\lqcd$ such that formalism like ChPT can be applied. Various technical issues arise when it is used in quarkonium physics~\cite{Ananthanarayan:2023gzw,Meng:2022ozq,Casalbuoni:1996pg,Brambilla:1999xf,Brambilla:2010cs,Brambilla:2004jw,Mannel:2020ups}. However, Lattice QCD methods have significantly developed over the years, and now precise predictions for the charm quarks are now available in the literature. We use experimental inputs as well as the results from the lattice QCD on the moments in the extraction of the $m_c$ in this section.\par In this section, we determine the $m_c$ from the vector moments and pseudoscalar moments using FOPT and RGSPT. As observed in section~\eqref{sec:RGmom}, these determinations from higher moments in the vector channel are very sensitive to scale variations, and the $\msbar$ definition in the condensate terms also causes trouble. The determination from the pseudoscalar channel does not suffer very much from these issues. 
\label{sec:mc_det}
\subsection{\texorpdfstring{$m_c$}{} determination using experimental inputs for the vector channel.}
For the vector channel, we use the experimental moments provided in Refs.~\cite{Chetyrkin:2017lif,Dehnadi:2011gc} in the $m_c$ determination. These moments are presented in Table~\eqref{tab:momentsV_charm}, and our results for the $m_c$ determination in the $\msbar$ scheme are presented in Table~\eqref{tab:mc_V_ms}.  A significant condensate contributions in the $m_c$ determination using FOPT is found for $\mathcal{M}^{V}_3$ and $\mathcal{M}^{V}_4$. This is caused by quark mass in the $\msbar$ scheme used in the condensate terms for FOPT as pointed out in Refs.~\cite{Chetyrkin:2010ic,Dehnadi:2011gc} and they use pole mass value in condensate to avoid such large contributions. However, this is not the case for the RGSPT determinations, which are also stable with respect to the scale variations. When we use the pole mass in the condensate, the $m_c$ determination using FOPT and RGSPT are presented in Table~\eqref{tab:mc_V_pole}. In this case, the situation is a little bit improved for FOPT, but scale dependence is still the major source of theoretical uncertainties. 

\begin{table}[H]
    \centering
    \begin{tabular}{|c|c|c|}
    \hline
        \textbf{Moments} &  \textbf{Ref.~\cite{Dehnadi:2011gc}}& \textbf{Ref.~\cite{Chetyrkin:2017lif}}\\\hline\hline
        $\mathcal{M}^{V,\text{exp.}}_1$&$2.121\pm0.036$ &$2.154\pm0.023$\\ \hline 
        $\mathcal{M}^{V,\text{exp.}}_2$&$1.478\pm0.028$ &$1.490\pm0.017$\\\hline
        $\mathcal{M}^{V,\text{exp.}}_3$&$1.302\pm0.027$ &$1.308\pm0.016$\\ \hline
        $\mathcal{M}^{V,\text{exp.}}_4$&$1.243\pm0.028$ &$1.248\pm0.016$\\ \hline
    \end{tabular}
    \caption{Moments for the vector channel for the charm case. These moments are in the units of $10^{-n}\GeV^{-2n}$.}
    \label{tab:momentsV_charm}
\end{table}

\begin{widetext}

\begin{table}[H]
    \centering
    \begin{tabular}{||c|c|c|c|c|c|c|c|c|c|c|c|c|c||}\hline\hline
    \text{}&\text{} &\multicolumn{6}{c|}{\textbf{FOPT}}&\multicolumn{6}{c|}{\textbf{RGSPT}}\\\cline{3-14}
    \textbf{Sources}&\textbf{Moments}&\text{}&\multicolumn{4}{c|}{Theo. Unc.}& \text{}&\text{}&\multicolumn{4}{c|}{Theo. Unc.}& \text{} \\
\cline{4-7}\cline{10-13}
\text{}&\text{}&$m_c(3\GeV)$&$\as$&$\mu$&n.p.&total&Exp. Unc.&$m_c(3\GeV)$&$\as$&$\mu$&n.p.&total&Exp. Unc.
\\\hline\hline
        \multirow{3}{4em}{ Ref.~\cite{Dehnadi:2011gc}}
               &$\mathcal{M}^{V}_1$&$1005.4( 13.9)$&3.2&7.6&0.2&8.3&11.2&$1000.2( 12.3)$&3.7&1.9&2.3&4.8&11.3\\\cline{2-14}
        \text{}&$\mathcal{M}^{V}_2$&$997.2( 19.8)$&4.7&11.3&14.4&18.9&6.1&$988.5( 9.2)$&5.4&1.6&3.5&6.6&6.3\\ \cline{2-14}
         \text{}&$\mathcal{M}^{V}_3$&$1022.1( 127.8)$&3.4&41.8&120.6&127.7&4.0&$983.4( 9.2)$&6.7&1.4&3.9&7.8&4.9\\ \cline{2-14}
         \text{}&$\mathcal{M}^{V}_4$&$1077.3( 113.6)$&1.0&100.5&52.9&113.6&2.8&$980.5( 8.9)$&7.7&0.9&1.8&8.0&3.9\\\hline\hline
       \multirow{3}{4em}{Ref.~\cite{Chetyrkin:2017lif}}
                &$\mathcal{M}^{V}_1$&$995.4( 10.8)$&3.3&7.6&0.0&8.3&6.9&$990.1( 8.5)$&3.6&2.0&2.3&4.8&7.0\\ \cline{2-14}
         \text{}&$\mathcal{M}^{V}_2$&$994.6( 19.5)$&4.7&11.3&14.7&19.1&3.7&$985.8( 7.7)$&5.4&1.6&3.5&6.6&3.8\\ \cline{2-14}
         \text{}&$\mathcal{M}^{V}_3$&$1021.3( 126.5)$&3.4&41.9&126.5&133.3&2.3&$982.3( 8.3)$&6.7&1.4&3.9&7.8&2.8\\ \cline{2-14}
         \text{}&$\mathcal{M}^{V}_4$&$1076.8( 113.8)$&1.0&100.7&52.9&113.8&1.6&$979.8( 8.3)$&7.7&0.9&1.8&8.0&2.2\\\hline\hline
    \end{tabular}
    \caption{$m_c$ determinations using FOPT and RGSPT in $\msbar$ scheme using experimental inputs from Table~\eqref{tab:momentsV_charm}. Results are in the units of $\MeV$ and the scale dependence is calculated for the energy range $\mu\in[1,4]\hs\GeV$.}
    \label{tab:mc_V_ms}
\end{table}
%
\begin{table}[H]
    \centering
    \begin{tabular}{||c|c|c|c|c|c|c|c|c|c|c|c|c|c||}\hline\hline
    \text{}&\text{} &\multicolumn{6}{c|}{\textbf{FOPT}}&\multicolumn{6}{c|}{\textbf{RGSPT}}\\\cline{3-14}
    \textbf{Sources}&\textbf{Moments}&\text{}&\multicolumn{4}{c|}{Theo. Unc.}& \text{}&\text{}&\multicolumn{4}{c|}{Theo. Unc.}& \text{} \\
\cline{4-7}\cline{10-13}
\text{}&\text{}&$m_c(3\GeV)$&$\as$&$\mu$&n.p.&total&Exp. Unc.&$m_c(3\GeV)$&$\as$&$\mu$&n.p.&total&Exp. Unc.
\\\hline\hline
        \multirow{3}{4em}{ Ref.~\cite{Dehnadi:2011gc}}
               &$\mathcal{M}^{V}_1$&$1004.8(13.7)$&3.3&7.1&1.4&7.9&11.2&$1000.9( 12.1)$&3.7&1.9&0.9&4.3&11.3\\\cline{2-14}
        \text{}&$\mathcal{M}^{V}_2$&$989.4( 9.2)$&5.4&3.5&2.0&6.7&6.3&$989.6 (8.6)$&5.5&1.8&1.0&5.8&6.3\\ \cline{2-14}
         \text{}&$\mathcal{M}^{V}_3$&$990.9( 13.5)$&5.4&10.9&2.5&12.6&4.7&$984.8( 8.7)$&6.8&2.4&1.0&7.3&4.8\\ \cline{2-14}
         \text{}&$\mathcal{M}^{V}_4$&$1014.5( 37.7)$&3.8&37.2&2.9&37.5&3.4&$980.9( 9.8)$&8.0&3.9&0.9&9.0&3.9\\\hline\hline
       \multirow{3}{4em}{Ref.~\cite{Chetyrkin:2017lif}}
                &$\mathcal{M}^{V}_1$&$995.1( 10.3)$&3.3&6.8&0.7&7.6&7.0&$990.8( 8.2)$&3.8&2.0&0.9&4.3&7.0\\ \cline{2-14}
         \text{}&$\mathcal{M}^{V}_2$&$987.3( 7.4)$&5.4&3.2&0.8&6.3&3.8&$986.9 (7.0)$&5.5&1.8&1.0&5.9&3.8\\ \cline{2-14}
         \text{}&$\mathcal{M}^{V}_3$&$990.7( 12.7)$&5.8&10.9&2.7&12.4&2.7&$983.7( 7.8)$&6.8&2.4&1.1&7.3&2.8\\ \cline{2-14}
         \text{}&$\mathcal{M}^{V}_4$&$1014.9( 37.0)$&3.8&36.7&0.8&36.9&2.0&$980.2( 9.3)$&8.1&3.9&1.0&9.0&2.2\\\hline\hline
    \end{tabular}
    \caption{$m_c$ determinations using FOPT and RGSPT  using experimental inputs from Table~\eqref{tab:momentsV_charm}. The pole mass of the charm quark is used as input in the non-perturbative condensate terms. Results are in the units of $\MeV$, and the scale dependence is calculated for the energy range $\mu\in[1,4]\hs\GeV$.}
    \label{tab:mc_V_pole}
\end{table}

\end{widetext}
\subsection{\texorpdfstring{$m_c$}{} determination using the lattice QCD inputs.}
The moments for the vector currents are obtained using the experimental data on hadrons from the $e^+ e^-$ collision. However, this is not the case for the pseudoscalar channel, which is not realized in nature but can be computed using the lattice QCD simulations. We use the results for the reduced moments ($R_n$), a dimensionless quantity is reliably calculable from the lattice QCD in the Refs.~\cite{HPQCD:2008kxl,McNeile:2010ji,Maezawa:2016vgv,Petreczky:2019ozv,Petreczky:2020tky}. The regular moments calculated in the perturbative QCD are related to these reduced moments by the following relations~\cite{Boito:2020lyp,Dehnadi:2015fra}:
\begin{align}
    \mathcal{M}_n^P=T^P_{n,0}\left(\frac{R_{2n+4}}{m_{\eta_c}}\right)^{2n}\,,
    \label{eq:reduced_mom}
\end{align}
and the results are collected in Table~\eqref{tab:momP_charm}. It should be noted that the reduced moments provided in the Refs.~\cite{HPQCD:2008kxl,McNeile:2010ji} are converted to regular moments using the current PDG~\cite{ParticleDataGroup:2022pth} value $m_{\eta_c}=2.9839\pm0.0004$ for the $\eta_c$-meson.\par
The $m_c$ determination using FOPT and RGSPT from the lattice QCD moments are presented in Table~\eqref{tab:mc_P_ms}. These determinations do not suffer issues from the condensate terms; the determinations from the first two moments are precise and close. The RGSPT determinations are even better for all three moments. Since the results from the pseudoscalar channel in the $\msbar$ scheme are good enough, we do not find it necessary to give our determinations using on-shell mass as input for the condensate terms. 
\begin{widetext}

\begin{table}[ht]
    \centering
    \begin{tabular}{|c|c|c|c|c|c|}\hline
         \textbf{Moments}& \textbf{Ref.~\cite{HPQCD:2008kxl}} &\textbf{Ref.~ \cite{McNeile:2010ji}} &\textbf{Ref.~ \cite{Maezawa:2016vgv}} & \textbf{Ref.~\cite{Petreczky:2019ozv}}& \textbf{Ref.~\cite{Petreczky:2020tky}} \\\hline\hline
         $\mathcal{M}^{P}_1$&$1.404\pm0.019$&$1.395\pm0.005$&$1.385\pm0.007$&$1.386\pm0.005$&$1.387\pm0.004$\\\hline
         $\mathcal{M}^{P}_2$&$1.359\pm0.041$&$1.365\pm0.012$&$1.345\pm0.032$&$1.349\pm0.012$&$1.344\pm0.010$\\\hline
         $\mathcal{M}^{P}_3$&$1.425\pm0.059$&$1.415\pm0.010$&$1.406\pm0.048$&$1.461\pm0.050$&$1.395\pm0.022$\\\hline
    \end{tabular}
    \caption{Pseudoscalar moment calculated from lattice QCD for the charm case. These $\mathcal{M}^{P}_n$ are in the units of $10^{-n}\GeV^{-2n}$.}
    \label{tab:momP_charm}
\end{table}
\begin{table}[ht]
    \centering
    \begin{tabular}{||c|c|c|c|c|c|c|c|c|c|c|c|c|c||}\hline\hline
    \text{}&\text{} &\multicolumn{6}{c|}{\textbf{FOPT}}&\multicolumn{6}{c|}{\textbf{RGSPT}}\\\cline{3-14}
    \textbf{Sources}&\textbf{Moments}&\text{}&\multicolumn{4}{c|}{Theo. Unc.}& \text{}&\text{}&\multicolumn{4}{c|}{Theo. Unc.}& \text{} \\
\cline{4-7}\cline{10-13}
\text{}&\text{}&$m_c(3\GeV)$&$\as$&$\mu$&n.p.&total&Exp. Unc.&$m_c(3\GeV)$&$\as$&$\mu$&n.p.&total&Exp. Unc.
\\\hline\hline
          \multirow{3}{4em}{ Ref.~\cite{HPQCD:2008kxl}}
               &$\mathcal{M}^{P}_1$&$983.6( 10.0)$&1.1&5.0&2.4&5.7&8.2&$989.3( 9.0)$&1.4&3.5&0.7&3.8&8.1\\\cline{2-14}
        \text{}&$\mathcal{M}^{P}_2$&$988.3 (12.5)$&1.7&6.8&3.6&12.5&9.8&$990.5( 11.4)$&1.5&5.8&0.9&6.0&9.7\\ \cline{2-14}
         \text{}&$\mathcal{M}^{P}_3$&$998.9( 29.9)$&2.2&26.6&10.5&28.6&8.5&$985.4( 11.6)$&3.2&6.4&2.0&7.4&9.0\\\hline\hline
          \multirow{3}{4em}{ Ref.~\cite{McNeile:2010ji}}
               &$\mathcal{M}^{P}_1$&$987.1( 6.1)$&1.1&5.0&2.3&5.5&2.4&$992.8( 4.5)$&1.4&3.5&3.5&0.7&2.3\\\cline{2-14}
        \text{}&$\mathcal{M}^{P}_2$&$986.9( 8.3)$&1.7&6.7&3.6&7.8&2.7&$989.1( 6.6)$&1.5&5.8&0.9&6.0&2.7\\ \cline{2-14}
         \text{}&$\mathcal{M}^{P}_3$&$1000.2( 28.6)$&2.2&26.5&10.4&28.6&1.4&$986.9( 7.6)$&3.2&6.4&2.0&7.4&1.5\\\hline\hline
          \multirow{3}{4em}{ Ref.~\cite{Maezawa:2016vgv} }
               &$\mathcal{M}^{P}_1$&$991.7( 6.4)$&1.1&4.9&2.2&5.5&3.2&$997.3( 5.0)$&1.4&3.5&0.7&3.8&3.2\\\cline{2-14}
        \text{}&$\mathcal{M}^{P}_2$&$991.5 (10.9)$&1.6&6.8&3.5&7.9&7.5&$993.6 (9.6)$&1.5&5.8&0.9&6.1&7.5\\ \cline{2-14}
         \text{}&$\mathcal{M}^{P}_3$&$1001.5( 29.4)$&2.2&26.5&10.3&28.5&7.1&$988.2( 10.5)$&3.2&6.4&2.0&7.4&7.5\\\hline\hline
        \multirow{3}{4em}{ Ref.~\cite{Petreczky:2019ozv}}
               &$\mathcal{M}^{P}_1$&$991.2( 6.0)$&1.1&4.9&2.3&5.5&2.4&$996.8( 4.5)$&1.4&3.5&0.7&3.8&2.3\\\cline{2-14}
        \text{}&$\mathcal{M}^{P}_2$&$990.5( 8.3)$&1.6&6.8&3.5&7.9&2.8&$992.7(6.6)$&1.5&5.8&0.9&6.0&2.7\\ \cline{2-14}
         \text{}&$\mathcal{M}^{P}_3$&$993.8( 29.7)$&2.2&26.7&10.9&28.9&7.0&$980.0( 10.5)$&3.3&6.3&2.0&7.4&7.4\\\hline\hline
       \multirow{3}{4em}{Ref.~\cite{Petreczky:2020tky}}
                &$\mathcal{M}^{P}_1$&$990.6( 5.9)$&1.1&4.9&2.3&5.5&1.9&$996.2( 4.2)$&1.4&3.5&0.7&3.8&1.9\\ \cline{2-14}
         \text{}&$\mathcal{M}^{P}_2$&$991.6( 8.2)$&1.6&3.5&7.9&2.3&9.8&$993.7(6.5)$&1.5&5.8&0.9&6.1&2.3\\ \cline{2-14}
         \text{}&$\mathcal{M}^{P}_3$&$1003.2( 28.6)$&2.2&26.5&10.2&28.6&3.2&$989.9( 8.2)$&3.2&6.4&2.0&7.4&3.4\\\hline\hline
    \end{tabular}
    \caption{$m_c$ determinations using FOPT and RGSPT  using experimental inputs from Table~\eqref{tab:momP_charm}. The pole mass of the charm quark is used as input in the non-perturbative condensate terms. Results are in the units of $\MeV$ and the scale dependence is calculated for the energy range $\mu\in[1,4]\hs\GeV$.}
    \label{tab:mc_P_ms}
\end{table}

\end{widetext}

\section{Bottom quark mass determination} \label{sec:mb_det}
The bottom quark is heavy, and to the best of our knowledge, there are no direct lattice data on the moments in the case of the bottom quark. We use the experimental information on the moments provided in Refs.~\cite{Kuhn:2007vp,Chetyrkin:2009fv,Dehnadi:2015fra} for the vector channel in the bottom quark mass determinations. These moments from different sources are tabulated in Table~\eqref{tab:momV_bottom}. The $m_b$ determination using FOPT and RGSPT in the $\msbar$ scheme are presented in Table~\eqref{tab:mb_ms_cond_ms}. Similar to the charm case, FOPT again suffers from large non-perturbative contributions and scale dependence. The uncertainty from the condensate term is alone $~70\%$ of the central value for the third and fourth moment. When bottom quark mass in the on-shell scheme is used as input, this issue is resolved, and our $m_b$ determinations are presented in Table~\eqref{tab:mb_ms_cond_pole}.

\begin{table}[ht]
    \centering
    \begin{tabular}{||c|c|c|c|c||}\hline\hline
         \textbf{Moments} & \textbf{Ref.~\cite{Dehnadi:2015fra}} &\textbf{Ref.~ \cite{Chetyrkin:2009fv}} &\textbf{Ref.~ \cite{Kuhn:2007vp}}\\\hline\hline
         $\mathcal{M}^{V,\text{exp.}}_1$&$4.526\pm0.112$&$4.592\pm0.031$&$4.601\pm0.043$\\\hline
         $\mathcal{M}^{V,\text{exp.}}_2$&$2.834\pm0.052$&$2.872\pm0.028$&$2.881\pm0.037$\\\hline
         $\mathcal{M}^{V,\text{exp.}}_3$&$2.338\pm0.036$&$2.362\pm0.026$&$2.370\pm0.034$\\\hline
         $\mathcal{M}^{V,\text{exp.}}_4$&$2.154\pm0.030$&$2.170\pm0.026$&$2.178\pm0.032$\\\hline
    \end{tabular}
    \caption{Vector moments from different sources used as input for the bottom quark mass determination. These $\mathcal{M}^{V}_n$ are in the units of $10^{-\left(2n+1\right)}\GeV^{-2n}$.}
    \label{tab:momV_bottom}
\end{table}

\begin{widetext}

\begin{table}[ht]
    \centering
    \begin{tabular}{||c|c|c|c|c|c|c|c|c|c|c|c|c|c||}\hline\hline
    \text{}&\text{} &\multicolumn{6}{c|}{\textbf{FOPT}}&\multicolumn{6}{c|}{\textbf{RGSPT}}\\\cline{3-14}
    \textbf{Sources}&\textbf{Moments}&\text{}&\multicolumn{4}{c|}{Theo. Unc.}& \text{}&\text{}&\multicolumn{4}{c|}{Theo. Unc.}& \text{} \\
\cline{4-7}\cline{10-13}
\text{}&\text{}&$m_b(10\GeV)$&$\as$&$\mu$&n.p.&total&Exp. Unc.&$m_b(10\hs\GeV)$&$\as$&$\mu$&n.p.&total&Exp. Unc.
\\\hline\hline
         \multirow{4}{4em}{ Ref.~\cite{Dehnadi:2015fra}}
          &$\mathcal{M}^{V}_1$&$3632.2( 53.4)$&3.2&5.4&0.0&6.3&53.0&$3631.6(53.1)$&3.2&0.7&0.0&3.3&53.0\\\cline{2-14}
        \text{}&$\mathcal{M}^{V}_2$&$3632.8( 21.0)$&5.2&5.3&0.0&7.4&19.7&$3633.1( 20.4)$&5.2&0.4&0.1&5.2&19.7\\ \cline{2-14}
         \text{}&$\mathcal{M}^{V}_3$&$3637.8( 2746.7)$&6.2&20.2&2746.6&2746.7&11.0&$3634.1( 12.8)$&6.5&0.3&0.2&6.5&11.1\\\cline{2-14}
         \text{}&$\mathcal{M}^{V}_4$&$3648.8( 2491.7)$&6.5&49.9&2491.2&2491.7&7.4&$3635.0( 10.5)$&7.4&0.4&0.2&7.4&7.5\\
         \hline\hline
        \multirow{4}{4em}{ Ref.~\cite{Chetyrkin:2009fv}}
               &$\mathcal{M}^{V}_1$&$3601.9(14.1)$&3.2&5.3&0.0&6.2&14.1&$3601.3 (14.5)$&3.2&0.7&0.0&3.3&14.1\\\cline{2-14}
        \text{}&$\mathcal{M}^{V}_2$&$3618.7( 12.7)$&5.2&5.2&0.0&7.4&10.4&$3619.0( 11.6)$&5.2&0.4&0.1&5.2&10.4\\ \cline{2-14}
         \text{}&$\mathcal{M}^{V}_3$&$3630.5( 2739.5)$&6.2&20.2&2739.4&2739.5&7.8&$3626.8( 10.2)$&6.5&0.3&0.2&6.5&7.9\\\cline{2-14}
         \text{}&$\mathcal{M}^{V}_4$&$3644.9( 2487.8)$&6.5&49.9&2487.3&2487.8&6.4&$3631.0( 9.8)$&7.4&0.4&0.2&7.4&6.5\\
         \hline\hline
       \multirow{4}{4em}{Ref.~\cite{Kuhn:2007vp}}
                &$\mathcal{M}^{V}_1$&$3597.8( 20.6)$&3.2&5.3&0.0&6.2&19.6&$3597.2( 19.9)$&3.2&0.7&0.0&3.3&19.6\\ \cline{2-14}
         \text{}&$\mathcal{M}^{V}_2$&$3615.4( 15.5)$&5.2&5.2&0.0&7.4&13.7&$3615.7( 14.6)$&5.2&0.4&0.1&5.2&13.7\\ \cline{2-14}
         \text{}&$\mathcal{M}^{V}_3$&$3628.1( 2737.1)$&6.2&20.3&2737.0&2737.1&10.2&$3624.4( 12.1)$&6.5&0.3&0.2&6.5&10.3\\ \cline{2-14}
         \text{}&$\mathcal{M}^{V}_4$&$3642.9( 2485.9)$&6.5&50.0&2485.3&2485.3&7.8&$3629.0(10.8)$&7.4&0.4&0.2&7.4&7.9\\\hline\hline
    \end{tabular}
    \caption{$m_b$ determinations using FOPT and RGSPT in the $\msbar$ scheme using experimental inputs from Table~\eqref{tab:momV_bottom}. Results are in the units of $\MeV$ and the scale dependence is calculated for the energy range $\mu\in[2,20]\hs\GeV$.}
    \label{tab:mb_ms_cond_ms}
\end{table}

\begin{table}[H]
    \centering
    \begin{tabular}{||c|c|c|c|c|c|c|c|c|c|c|c|c|c||}\hline\hline
    \text{}&\text{} &\multicolumn{6}{c|}{\textbf{FOPT}}&\multicolumn{6}{c|}{\textbf{RGSPT}}\\\cline{3-14}
    \textbf{Sources}&\textbf{Moments}&\text{}&\multicolumn{4}{c|}{Theo. Unc.}& \text{}&\text{}&\multicolumn{4}{c|}{Theo. Unc.}& \text{} \\
\cline{4-7}\cline{10-13}
\text{}&\text{}&$m_b(10\GeV)$&$\as$&$\mu$&n.p.&total&Exp. Unc.&$m_b(10\hs\GeV)$&$\as$&$\mu$&n.p.&total&Exp. Unc.
\\\hline\hline
         \multirow{4}{4em}{ Ref.~\cite{Dehnadi:2015fra}}
          &$\mathcal{M}^{V}_1$&$3632.2( 53.4)$&3.2&5.4&0.0&6.3&53.0&$3631.6( 53.0)$&3.2&0.7&0.0&3.3&53.0\\\cline{2-14}
        \text{}&$\mathcal{M}^{V}_2$&$3632.8( 21.0)$&5.2&5.3&0.0&7.4&19.7&$3633.2( 20.4)$&5.2&0.4&0.0&5.2&19.7\\ \cline{2-14}
         \text{}&$\mathcal{M}^{V}_3$&$3637.6( 23.2)$&6.2&19.5&0.0&20.5&11.0&$3634.2( 12.8)$&6.5&0.3&0.0&6.5&11.1\\\cline{2-14}
         \text{}&$\mathcal{M}^{V}_4$&$3648.3( 48.8)$&6.5&47.8&0.1&48.3&7.4&$3635.1( 10.5)$&7.4&0.4&0.0&7.4&7.5\\
         \hline\hline
        \multirow{4}{4em}{ Ref.~\cite{Chetyrkin:2009fv}}
               &$\mathcal{M}^{V}_1$&$3601.9( 15.5)$&3.2&5.3&0.0&6.2&14.1&$3601.3( 14.5)$&3.2&0.7&0.0&3.3&14.1\\\cline{2-14}
        \text{}&$\mathcal{M}^{V}_2$&$3618.7( 12.7)$&5.2&5.2&0.0&7.4&10.4&$3619.1( 11.6)$&5.2&0.4&0.0&5.2&10.4\\ \cline{2-14}
         \text{}&$\mathcal{M}^{V}_3$&$3630.4( 22.0)$&6.2&19.6&0.0&20.5&7.8&$3626.9( 10.2)$&6.5&0.3&0.0&6.5&7.9\\\cline{2-14}
         \text{}&$\mathcal{M}^{V}_4$&$3644.3( 48.7)$&6.5&47.9&0.1&48.3&6.4&$3631.1( 9.8)$&7.4&0.4&0.0&7.4&6.5\\
         \hline\hline
       \multirow{4}{4em}{Ref.~\cite{Kuhn:2007vp}}
                &$\mathcal{M}^{V}_1$&$3597.8( 20.6)$&3.2&5.3&0.0&6.2&19.6&$ 3597.3( 19.9)$&3.2&0.7&0.0&3.3&19.6\\ \cline{2-14}
         \text{}&$\mathcal{M}^{V}_2$&$3615.4( 15.5)$&5.2&5.2&0.0&7.3&13.7&$3615.8( 14.6)$&5.2&0.4&0.0&5.2&13.7\\ \cline{2-14}
         \text{}&$\mathcal{M}^{V}_3$&$3628.0( 22.9)$&6.2&19.6&0.0&20.5&10.2&$3624.5( 12.1)$&6.5&0.3&0.0&6.5&10.3\\ \cline{2-14}
         \text{}&$\mathcal{M}^{V}_4$&$3642.4( 48.9)$&6.5&47.9&0.1&48.3&7.8&$3629.1( 10.8)$&7.4&0.4&0.0&7.4&7.9\\\hline\hline
    \end{tabular}
    \caption{$m_b$ determinations using FOPT and RGSPT  using experimental inputs from Table~\eqref{tab:momV_bottom}. The pole mass of the bottom quark is used as input in the non-perturbative condensate terms. Results are in the units of $\MeV$, and the scale dependence is calculated for the energy range $\mu\in[2,20]\hs\GeV$.}
    \label{tab:mb_ms_cond_pole}
\end{table}

\end{widetext}

\section{\texorpdfstring{$\as$}{} determination} \label{sec:as_det}
For the $\as$ determination, instead of the $\mathcal{M}^{X}_n$, we use the dimensionless quantities such as $\mathcal{M}^{P}_0$ and the ratio of moments $\mathcal{R}^{X}_n$ defined Eq.~\eqref{eq:Def_R}. Theoretical expressions are sensitive to the $\as$, and the quark mass dependence appears at NNLO via running logarithms. These quantities are very important observables for $\as$ determination. These ratios can also be calculated from the lattice QCD for the charm case in the pseudoscalar channel. We do not get any reliable determinations of the $\as$ for the bottom moments in the vector channel. Therefore, our determinations are only based on the charmonium sum rules.\par 
We use the ratios of the moments for the vector channel provided in Ref.~\cite{Boito:2019pqp,Boito:2020lyp} in the $\as$ determinations. For the pseudoscalar channel, we use results on the moments and from Refs.~\cite{HPQCD:2008kxl,McNeile:2010ji,Maezawa:2016vgv,Nakayama:2016atf,Petreczky:2019ozv,Petreczky:2020tky}. \par 
It should be noted that the $\as$ determination in this section is first performed at charm quark mass scale $m_c(m_c)=1.27\pm0.2\hs\GeV$ and then evolved to boson mass scale ($M_Z=91.18\hs\GeV$) by performing the matching and decoupling at the bottom quark mass scale using $m_b(m_b)=4.18\hs\GeV$ in the $\msbar$ scheme~\cite{ParticleDataGroup:2022pth}. We have used the REvolver package to perform the running and decoupling once $\as(\overline{m}_c)$ is obtained.\par Another technical point is the evolution of quark mass when uncertainties coming from scale variation are calculated. For this, we have taken our $x(\overline{m}_c)=\as(\overline{m}_c)/\pi$ determination as input and numerically solved for $x(q)$ at different scale, $q$, using relation:
\begin{align}
    m_c(q)=\overline{m}_c \int_{x(\overline{m}_c)}^{x(q)} \hs d\hs x\hs e^{\left(\frac{\gamma(x)}{\beta(x)}\right)}\,,
\end{align}
where $\gamma(x)$ and $\beta(x)$ are the five-loop quark mass anomalous dimension and QCD beta function. 

\subsection{\texorpdfstring{$\as$}{} from the vector channel.}
For the vector channel moments, one needs the experimental information about the resonances, and additional continuum contributions are modeled using the theoretical expression for the hadronic $R-$ratio for $e^+e^-$ in Eq.~\eqref{eq:Rratio}. We do not calculate these moments in this section, instead using very recent results provided in Ref.~\cite{Boito:2019pqp,Boito:2020lyp}. We have collected experimental inputs in Table~\eqref{tab:RV_charm}. We have tabulated our determinations in the $\msbar$ scheme in Table~\eqref{tab:as_V_ms}. Contrary to the previous sections, the effects of the non-perturbative terms in the $\msbar$ scheme are the same for $\as$ determinations from FOPT and RGSPT. Even though RGSPT has significant control over the renormalization scale uncertainties, uncertainties arising from the non-perturbative contributions dominate in the higher moments. \par 
We also perform $\as$ determinations to control these uncertainties using the numerical value of the on-shell mass for the charm quark in the condensate terms. The results obtained are presented in Table~\eqref{tab:as_V_pole}. We have found that using the on-shell mass in the non-perturbative term significantly improves our $\as$ determination. It is remarkable to note that the theoretical uncertainties are of similar size to the experimental ones for RGSPT.\par
\begin{figure}[ht]
\centering
		\includegraphics[width=.45\textwidth]{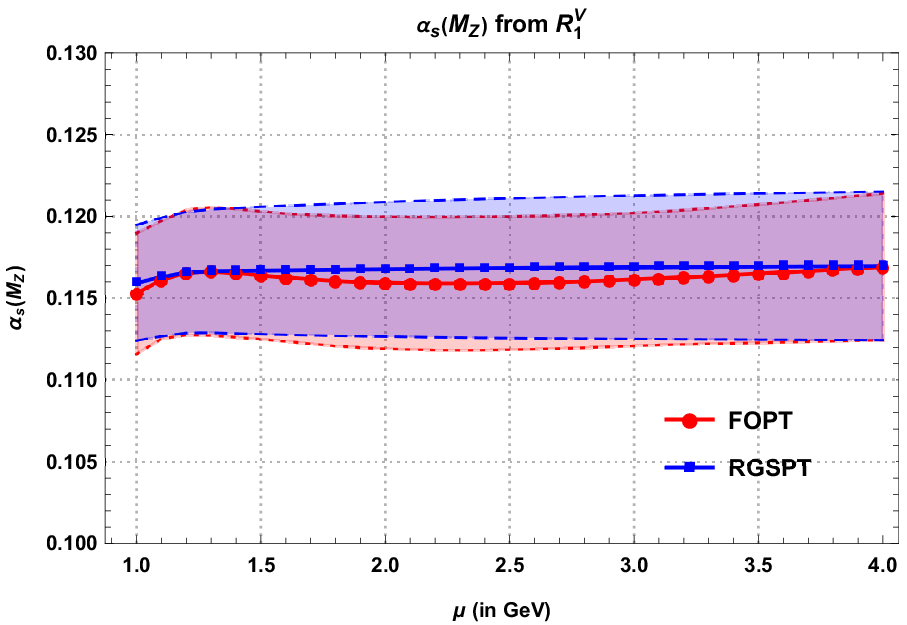}
		\includegraphics[width=.45\textwidth]{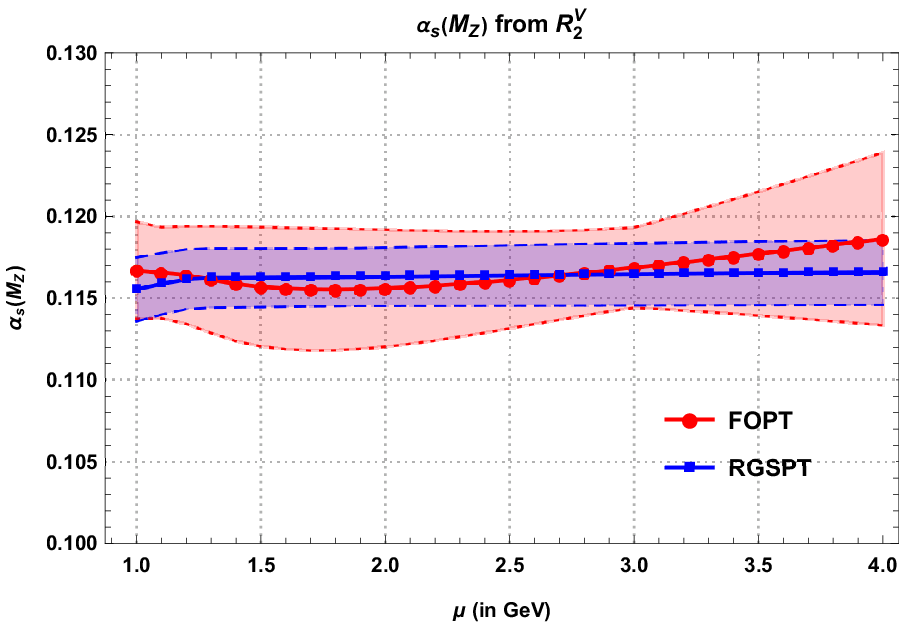}
	\includegraphics[width=.45\textwidth]{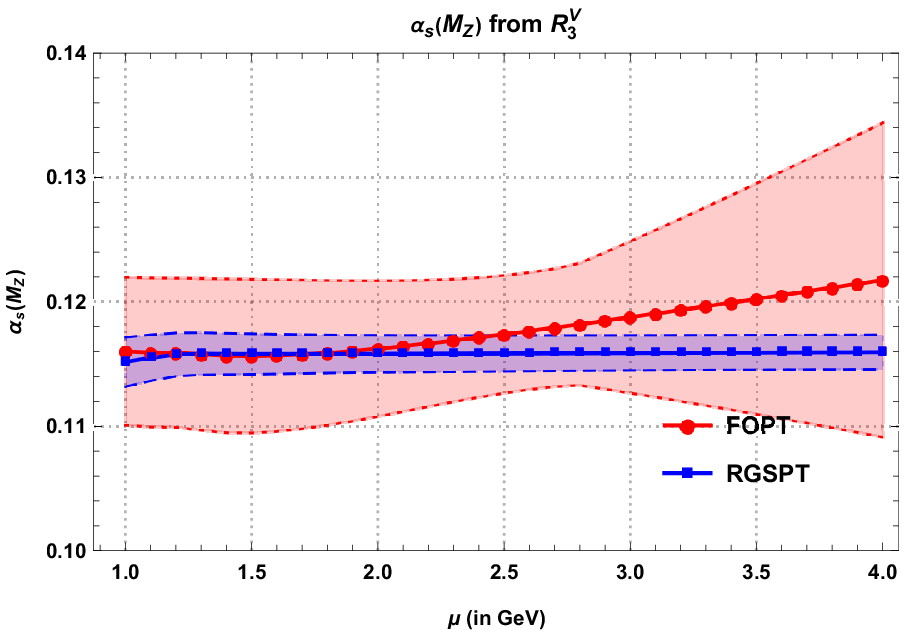}
\caption{\label{fig:asmz_V} $\as$ determination from the $R_n^{V}$ using $\msbar$ value of the charm quark mass as input in the condensate term. The bands represent the total uncertainty in the determinations.}
\end{figure}
\begin{figure}[ht]
\centering
		\includegraphics[width=.45\textwidth]{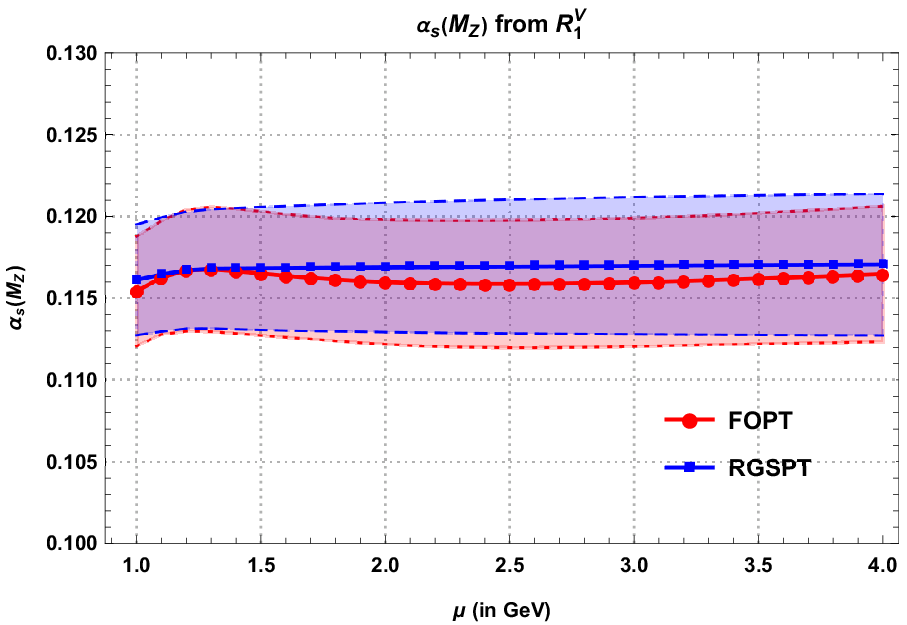}
		\includegraphics[width=.45\textwidth]{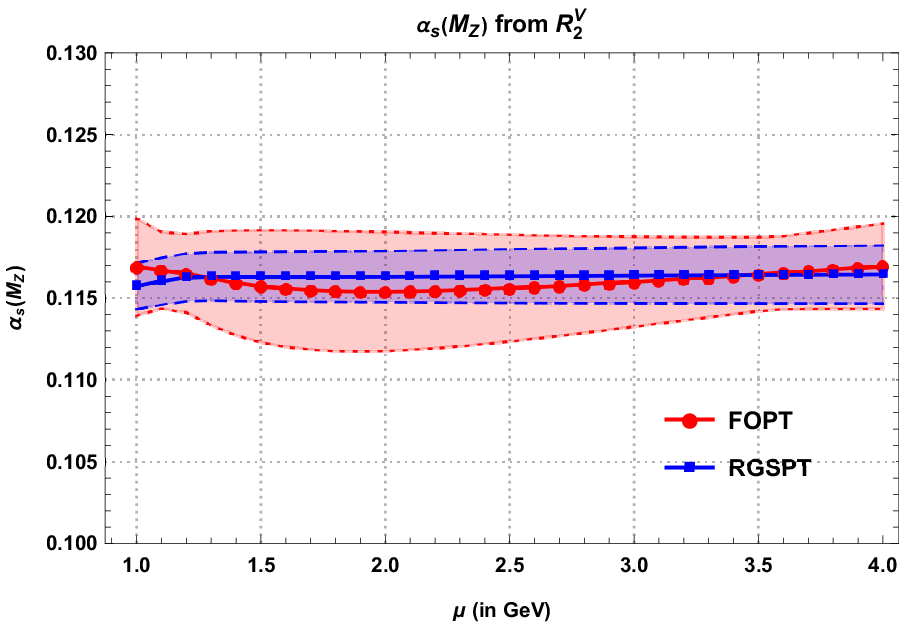}
	\includegraphics[width=.45\textwidth]{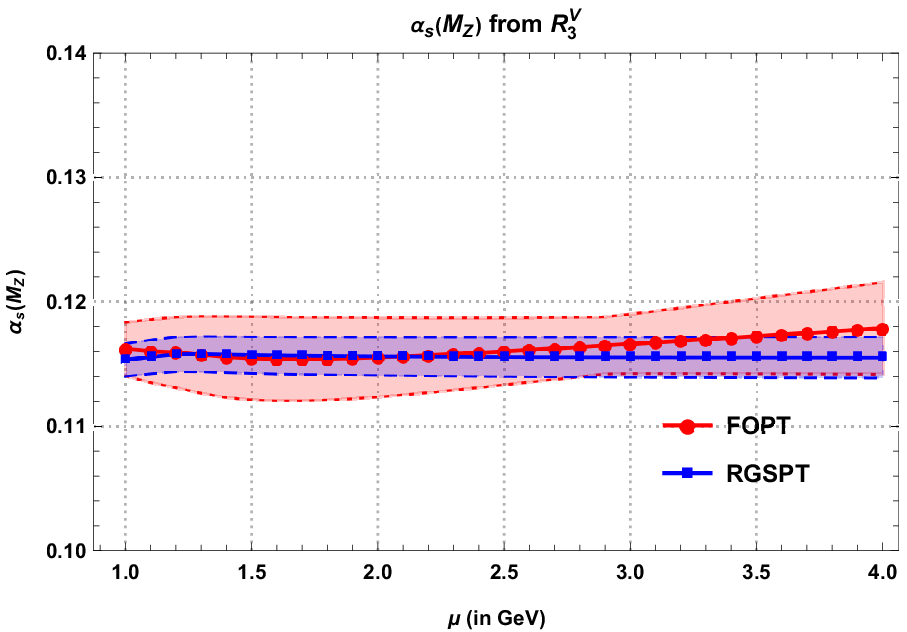}
\caption{\label{fig:asmz_V_pole} $\as$ determination from the $R_n^{V}$ using the on-shell value of the charm quark mass as input in the condensate term. The bands represent the total uncertainty in the determinations.}
\end{figure}

We can also notice that our determination in Tables~\eqref{tab:as_V_ms},\eqref{tab:as_V_pole} have the same central value for both RGSPT and FOPT. Due to the choice of scale $\mu=\overline{m}_c$, FOPT and RGSPT have the same expressions for the moments. Different scale choices result in different central values and uncertainties. This behavior can be seen in Fig.~\eqref{fig:asmz_V} and Fig.~\eqref{fig:asmz_V_pole} for the two scenarios considered above for the treatment of the non-perturbative terms. These plots also show remarkable stability in the $\as$ determinations from the RGSPT.  
\begin{table}[H]
    \centering
    \begin{tabular}{||c|c||}\hline\hline
         $n$& $\mathcal{R}^{V}_n$ \\\hline\hline
         1&$1.770\pm0.017$\\\hline
         2&$1.1173\pm0.0023$\\\hline
         3&$1.03536\pm0.00084$\\\hline\hline
    \end{tabular}
    \caption{Ratio of the experimental moments for the vector current obtained using the current PDG~\cite{ParticleDataGroup:2022pth} value of the $\as^{\left(5\right)}(M_Z)=0.1179\pm 0.0009$.}
    \label{tab:RV_charm}
\end{table}
\begin{widetext}

\begin{table}[H]
    \centering
    \begin{tabular}{||c|c|c|c|c|c|c|c|c|c|c|c|c||}
    \hline\hline
    \text{}&\multicolumn{6}{c|}{\textbf{FOPT}}&\multicolumn{6}{c|}{\textbf{RGSPT}}\\\cline{2-13}
    \textbf{Moment}&\text{}&\multicolumn{4}{c|}{Theo. Unc.}&\text{}&\text{}&\multicolumn{4}{c|}{Theo. Unc.}& \text{} \\
		\cline{3-6}\cline{9-12}
\text{}&$\as(M_Z)$&$m_c$&$\mu$&n.p.&total& Exp. Unc.&$\as(M_Z)$&$m_c$&$\mu$&n.p.&total& Exp. Unc. \\ \hline\hline
  $\mathcal{R}^{V}_1$&$0.1167(39)$&3&13&8&16&36&$0.1167(38)$&3&7&8&11&36\\\hline
  $\mathcal{R}^{V}_2$&$0.1163(31)$&4&27&12&29&11&$0.1163(18)$&3&8&12&15&11\\\hline
  $\mathcal{R}^{V}_3$&$0.1159(60)$&4&58&14&60&5&$0.1159(17)$&3&6&14&16&5\\\hline\hline
    \end{tabular}
    \caption{$\as$ determination using FOPT and RGSPT for the vector channel and sources of uncertainties from different sources. The scale dependence is calculated for the energy range $\mu\in[1,4]\hs\GeV$. The $\msbar$ scheme value for the charm quark is used in condensate terms.}
    \label{tab:as_V_ms}
\end{table}

\begin{table}[H]
    \centering
    \begin{tabular}{||c|c|c|c|c|c|c|c|c|c|c|c|c||}
    \hline\hline
    \text{}&\multicolumn{6}{c|}{\textbf{FOPT}}&\multicolumn{6}{c|}{\textbf{RGSPT}}\\\cline{2-13}
    \textbf{Moment}&\text{}&\multicolumn{4}{c|}{Theo. Unc.}&\text{}&\text{}&\multicolumn{4}{c|}{Theo. Unc.}& \text{} \\
		\cline{3-6}\cline{9-12}
\text{}&$\as(M_Z)$&$m_c$&$\mu$&n.p.&total& Exp. Unc.&$\as(M_Z)$&$m_c$&$\mu$&n.p.&total& Exp. Unc. \\ \hline\hline
 $\mathcal{R}^{V}_1$&$0.1169(38)$&3&13&5&15&35&$0.1169(36)$&2&6&5&8&35\\\hline
  $\mathcal{R}^{V}_2$&$0.1164(28)$&4&24&9&26&10&$0.1164(15)$&2&5&6&8&10\\\hline
  $\mathcal{R}^{V}_3$&$0.1159(30)$&3&27&13&30&5&$0.1159(14)$&2&2&6&6&5\\\hline\hline
    \end{tabular}
    \caption{$\as$ determination using FOPT and RGSPT for the vector channel and sources of uncertainties from different sources. The scale dependence is calculated for the energy range $\mu\in[1,4]\hs\GeV$. The on-shell mass of the charm quark is used in condensate terms.}
    \label{tab:as_V_pole}
\end{table}
\begin{table}[H]
    \centering
    \begin{tabular}{|c|c|c|c|c|c|c|}\hline
        \textbf{ Moments} & \textbf{Ref.~\cite{HPQCD:2008kxl}}& \textbf{Ref.~\cite{McNeile:2010ji}} &\textbf{Ref.~ \cite{Maezawa:2016vgv} }&\textbf{Ref.~\cite{Nakayama:2016atf}} & \textbf{Ref.~\cite{Petreczky:2019ozv}}& \textbf{Ref.~\cite{Petreczky:2020tky}} \\\hline\hline
         $\mathcal{M}^{P}_0$&$1.708\pm0.007$    &$1.708\pm0.005$    &$-$                &$1.699\pm0.008$    &$1.705\pm0.005$    &$1.7037\pm0.0027$\\\hline
         $\mathcal{R}^{P}_1$&$1.197\pm0.004$    &$-$                &$1.188\pm0.004$    &$1.199\pm0.004$    &$1.1886\pm0.013$   &$1.1881\pm0.0007$\\\hline
         $\mathcal{R}^{P}_2$&$1.033\pm0.004$    &$-$                &$1.0341\pm0.0018$  &$1.0344\pm0.0013$  &$1.0324\pm0.0016$  &$-$\\\hline
    \end{tabular}
    \caption{Pseudoscalar moment calculated from lattice QCD for the charm case. These $\mathcal{M}^{P}_n$ are in the units of $10^{-n}\GeV^{-2n}$.}
    \label{tab:momP_as_charm}
\end{table}

\begin{table}[H]
    \centering
    \begin{tabular}{|c|c|c|c|c|c|c|c|c|c|c|c|c|c|}\hline
    \text{}&\text{} &\multicolumn{6}{c|}{\textbf{FOPT}}&\multicolumn{6}{c|}{\textbf{RGSPT}}\\\cline{3-14}
    \textbf{Sources}&\textbf{Moments}&\text{}&\multicolumn{4}{c|}{Theo. Unc.}& \text{}&\text{}&\multicolumn{4}{c|}{Theo. Unc.}& \text{} \\
\cline{4-7}\cline{10-13}
\text{}&\text{}&$\as(M_Z)$&$m_c$&$\mu$&n.p.&total&Exp. Unc.&$\as(M_Z)$&$m_c$&$\mu$&n.p.&total&Exp. Unc.
\\\hline\hline
         \multirow{3}{4em}{ Ref.~\cite{HPQCD:2008kxl}}&$\mathcal{M}^{P}_0$&0.1172(20)&3&19&3&19&6&0.1172(8)&3&3&3&5&6\\
         \cline{2-14}
         \text{}&$\mathcal{R}^{P}_1$&0.1182(43)&4&42&5&43&6&0.1181(15)&3&12&5&13&6\\
         \cline{2-14}\text{}&$\mathcal{R}^{P}_2$&0.1150(53)&4&50&9&51&15&0.1149(18)&3&7&9&11&15\\\hline\hline
         \multirow{1}{4em}{  Ref.~\cite{McNeile:2010ji} }&$\mathcal{M}^{P}_0$&0.1172(20)&3&19&3&19&5&0.1172(7)&3&3&3&5&5\\\hline\hline
         \multirow{2}{4em}{ Ref.~\cite{Maezawa:2016vgv}}&$\mathcal{M}^{P}_0$&0.1168(48)&3&8&6&47&7&0.1168(13)&3&9&6&11&7\\
         \cline{2-14}
         \text{}&$\mathcal{R}^{P}_1$&0.1152(51)&4&50&8&50&6&0.1152(13)&3&7&8&11&6\\\hline\hline
        \multirow{3}{4em}{ Ref.~\cite{Nakayama:2016atf}}&$\mathcal{M}^{P}_0$&0.1164(20)&3&18&4&19&7&0.1164(9)&3&3&4&5&7\\
          \cline{2-14}\text{}&$\mathcal{R}^{P}_1$&0.1182(43)&4&42&5&43&6&0.1184(15)&3&13&5&14&6\\
         \cline{2-14}
         \text{}&$\mathcal{R}^{P}_2$&0.1153(50)&4&49&8&50&5&0.1153(12)&3&7&8&7&5\\\hline\hline
       \multirow{3}{4em}{Ref.~\cite{Petreczky:2019ozv}}&$\mathcal{M}^{P}_0$&0.1169(20)&3&19&3&19&5&0.1169(7)&3&13&3&5&5\\
         \cline{2-14} \text{}&$\mathcal{R}^{P}_1$&0.1169(47)&4&47&6&47&2&0.1169(12)&3&10&6&12&2\\
         \cline{2-14}
         \text{}&$\mathcal{R}^{P}_2$&0.1146(53)&4&52&9&53&6&0.1146(13)&3&6&9&11&6\\\hline\hline
       \multirow{2}{4em}{ Ref.~\cite{Petreczky:2020tky}}&$\mathcal{M}^{P}_0$&0.1168(19)&3&19&3&19&2&0.1168(13)&3&9&6&11&7\\
         \cline{2-14}
         \text{}&$\mathcal{R}^{P}_1$&0.1168(47)&4&47&6&47&1&0.1168(12)&3&1&6&11&1\\\hline
    \end{tabular}
    \caption{$\as$ determination from the pseudoscalar channel in the $\msbar$ scheme from various sources as input from Table~\eqref{tab:momP_as_charm}.}
    \label{tab:as_P_ms_charm}
\end{table}

\end{widetext}


\subsection{\texorpdfstring{$\as$}{} determination using the lattice QCD data}
The lattice QCD is the only source where the simulations can calculate pseudoscalar moments, and then the results are extrapolated to the continuum limits. Their results have their limits as the dimensionless moments and their ratios $\mathcal{R}^P_{n}$ for the higher moments suffer from the lattice artifacts. Despite these limitations, lower moments can still be used in the $\as$ determination. The lattice QCD simulations calculate the reduced moments $R_{2n+4}$, which will be transformed into the regular moments from the perturbative QCD. In addition to Eq.~\eqref{eq:reduced_mom}, we also need zeroth moment and dimensionless $\mathcal{R}_n^P$ using Eq.~\eqref{eq:Def_R} are obtained as~\cite{Boito:2020lyp,Dehnadi:2015fra}:
\begin{equation}
\begin{aligned}
\mathcal{M}_0^P&=T^P_{0,0} R_4\,,\\
  \mathcal{R}_n^P&=\frac{\left(T^P_{n,0}\right)^{1/n}}{\left(T^P_{n+1,0}\right)^{1/\left(n+1\right)}} \left(\frac{R_{2n+4}}{R_{2n+6}}\right)^2\,.
\end{aligned}
\label{eq:def_Rn}
\end{equation}
Using above relations, the results of Refs.~\cite{HPQCD:2008kxl,McNeile:2010ji,Maezawa:2016vgv,Petreczky:2019ozv,Petreczky:2020tky} are tabulated in Table~\eqref{tab:momP_as_charm}. We use these results in the $\as$ determination from the pseudoscalar moments, and the result in the $\msbar$ scheme are presented in Table~\eqref{tab:as_P_ms_charm}. The non-perturbative effects are under control since only the first few moments are used as input. They do not cause any issues in our determination, so there is no need to reiterate the exercise using the on-shell mass for the charm quark in condensate terms. Again, our determinations from the FOPT are dominated by theoretical uncertainties, especially the scale variation, but RGSPT gives very stable results. Interestingly, the determinations of $\as$ from $\mathcal{R}^{P}_2$ give smaller central values than the other moments considered. 
\section{Summary and Conclusion}\label{sec:summary}
In section~\eqref{sec:formulas}, the perturbative quantities related to the low energy moments of the current correlators are RG improved using RGSPT in section~\eqref{sec:RGmom}. The scale dependence of these RG invariant quantities using FOPT and RGSPT are plotted in Figs.~\eqref{fig:MomV_c}, \eqref{fig:MomP_c}, \eqref{fig:Mom_b}. It is evident from these plots that a more precise determination of the $m_c$, $m_b$, and $\as$ can be obtained.\par In section~\eqref{sec:mc_det}, the determination of the $m_c$ is performed using experimental vector and lattice pseudoscalar moments. The $m_c$ determinations using FOPT from the vector moments suffer from large uncertainties originating from the non-perturbative terms. This problem is not encountered with RGSPT; results are presented in Table~\eqref{tab:mc_V_ms}. To improve FOPT determination, we take quark mass in the on-shell scheme as input for the non-perturbative terms. This choice leads to improved determinations and results in Ref.~\eqref{tab:mc_V_pole}. The non-perturbative effects are not as problematic for the pseudoscalar moments as for the vector case. Slightly more precise values are obtained from the lattice moments than vector moments. These results are presented in Table~\eqref{tab:mc_P_ms}.\par
In section~\eqref{sec:mb_det}, the $m_b$ determination only using the vector moments is performed using FOPT and RGSPT. The condensate terms are more troubling in this case than in the $m_c$ determination. For the third and fourth moments, the determinations using FOPT suffer largely from the uncertainty from non-perturbative terms, which are nearly $70\%$ of the central value. These results are presented in Table~\eqref{tab:mb_ms_cond_ms}. When the numerical value of the bottom quark mass in the on-shell scheme is taken as input in the non-perturbative term, this problem for FOPT determination disappears. These results are presented in Table~\eqref{tab:mb_ms_cond_pole}.\par
In section~\eqref{sec:as_det}, the $\as$ determination is performed using dimensionless moments and ratios of the moments for the charm vector and pseudoscalar moments. The values obtained using the vector moments have similar issues from the non-perturbative terms as in the case of the $m_c$ and $m_b$ determination. These are again solved using the on-shell charm quark mass. These results are presented in Tables~\eqref{tab:as_V_ms} and \eqref{tab:as_V_pole}. Determinations using the pseudoscalar moments are presented in Table~\eqref{tab:as_P_ms_charm}.\par
In addition, it is worth mentioning that the RGSPT can also be used to calculate the continuum contributions to experimental moments where electromagnetic R-ratio is taken as input. In Ref.~\cite{AlamKhan:2023dms}, a significant reduction in the theoretical uncertainties originating from renormalization scale variation and truncation of the perturbation series is obtained for R-ratio. As an application of the method developed, light quark masses determined in Ref.~\cite{AlamKhan:2023ili} are more precise compared to the FOPT. These results can also be used for the method used in Ref.~\cite{Erler:2016atg,Erler:2022mzd} for the continuum contributions.\par
Now, we turn to the final values for the $m_c$, $m_b$, and $\as$ determination. We take the most precise values obtained in this article. Interestingly, all of them are obtained using RGSPT and lattice inputs except for the bottom quark mass, for which no lattice moments are available. For the charm mass, we give our final determination that is obtained from Table~\eqref{tab:mc_P_ms} using Ref.~\cite{Petreczky:2020tky} as:
\begin{align}
    m_c(3\GeV)&=0.9962(42)\hs\GeV\,,\\
    \implies m_c(m_c)&=1.2811(38)\hs\GeV\,.
\end{align}
\par
For the bottom quark mass, we take the most precise value obtained in Table~\eqref{tab:mb_ms_cond_pole} from Ref.~\cite{Chetyrkin:2009fv} as:
\begin{align}
    m_b(10\GeV)&=3.6311(98)\hs\GeV\,,\\
    \implies m_b(m_b)&=4.1743(95)\hs\GeV\,.
\end{align}
We have two most precise determinations for strong coupling constant in Table~\eqref{tab:as_P_ms_charm} from Refs.~\cite{McNeile:2010ji,Petreczky:2019ozv}. We average out these values and obtain the final determination:
\begin{align}
    \as(M_Z)&=\{0.1172(7),0.1169(7)\}\,,\\
    \implies  \as(M_Z)&=0.1171(7)\,.
\end{align}
These values are in full agreement with the current PDG~\cite{ParticleDataGroup:2022pth} values which read:
\begin{align}
    \as(M_Z)&=0.1179(9)\,,\\
     m_c(m_c)&=1.27\pm0.02\hs\GeV\,,\\
     m_b(m_b)&=4.18\pm0.03\hs\GeV\,.
\end{align}

\section*{Acknowledgment}
We thank Prof. B. Ananthanarayan and Mr. Aadarsh Singh for carefully reading the manuscript and for their valuable comments. The author is also thankful to Prof. Apoorva Patel for the financial support. The author is also supported by a scholarship from the Ministry of Human Resource Development (MHRD), Govt. of India. We are also thankful to Dr. Greynat for clarifying remarks about Refs.~ \cite{Greynat:2010kx,Greynat:2011zp}. This work is a part of the author's Ph.D. thesis.

\appendix

\section{Perturbative coefficients of the moments} \label{app:pert_coef}
 The perturbative coefficients of the vector and pseudoscalar moments for the charm case are presented in Table~\eqref{tab:coef_V_charm} and Table~\eqref{tab:coef_P_charm}, respectively. For the bottom case, only vector moments are used and their coefficients are presented in Table~\eqref{tab:coef_V_bottom}. 
\begin{table}[H]
    \centering
    \begin{tabular}{||c||c|c|c|c|c|c||}
    \hline\hline
        Moments& $T^V_{0,0}$& $T^V_{1,0}$& $T^V_{2,0}$& $T^V_{3,0}$& $T^{V,\text{n.p.}}_{0,0}$& $T^{V,\text{n.p.}}_{1,0}$\\\hline 
        $\mathcal{M}^V_1$ &1.067& 2.555& 2.497& -5.640&-0.251& -0.235\\\hline 
        $\mathcal{M}^V_2$ &0.457&1.110& 2.777& -3.494 & -0.104 & 0.051\\\hline
        $\mathcal{M}^V_3$ &0.271 & 0.519 & 1.639 & -2.840 & -0.038 & 0.077 \\\hline
        $\mathcal{M}^V_4$ &0.185 & 0.203 & 0.796 & -3.348 & -0.013 & 0.047\\\hline\hline
    \end{tabular}
    \caption{The perturbative coefficients of the vector moments for the charm case in the $\msbar$ scheme.}
    \label{tab:coef_V_charm}
\end{table}
\begin{table}[H]
    \centering
    \begin{tabular}{||c||c|c|c|c|c|c||}
    \hline\hline
        Moments& $T^P_{0,0}$& $T^P_{1,0}$& $T^P_{2,0}$& $T^P_{3,0}$&$T^{P,\text{n.p.}}_{0,0}$& $T^{P,\text{n.p.}}_{1,0}$\\\hline 
         $\mathcal{M}^P_0$ & 0.333& 0.778 & 0.183 & -1.806 & 0.877 & -0.045\\\hline 
        $\mathcal{M}^P_1$ & 0.133 & 0.516 & 1.881 & 1.515 & 0.125 & -0.391\\\hline 
       $\mathcal{M}^P_2$ & 0.076 & 0.303 & 1.567 & 3.711 & 0.0 & -0.142\\\hline 
       $\mathcal{M}^P_3$ & 0.051 & 0.178 & 1.138 & 3.583 & -0.009 & -0.022\\\hline\hline
    \end{tabular}
    \caption{The perturbative coefficients of the pseudoscalar moments for the charm case in the $\msbar$ scheme.}
    \label{tab:coef_P_charm}
\end{table}

\begin{table}[H]
    \centering
    \begin{tabular}{||c||c|c|c|c|c|c||}
    \hline\hline
        Moments& $T^V_{0,0}$& $T^V_{1,0}$& $T^V_{2,0}$& $T^V_{3,0}$&$T^{V,\text{n.p.}}_{0,0}$& $T^{V,\text{n.p.}}_{1,0}$\\\hline 
       $\mathcal{M}^V_1$ & 0.267 & 0.639 & 0.790 & -1.941 & -0.110 & -0.271\\\hline 
        $\mathcal{M}^V_2$ & 0.114 & 0.277 & 0.808 & -0.661 & -0.063 & -0.076 \\\hline 
      $\mathcal{M}^V_3$ & 0.068 & 0.130 & 0.517 & -0.294 & -0.026 & 0.005\\\hline 
     $\mathcal{M}^V_4$ &  0.046 &  0.051 & 0.305 & -0.346 & -0.009 & 0.017\\\hline \hline 
    \end{tabular}
    \caption{The perturbative coefficients of the vector moments for the bottom case in the $\msbar$ scheme.}
    \label{tab:coef_V_bottom}
\end{table}

\begin{widetext}

\section{Solution to summed coefficients}
\label{app:RGcoefs}
The solution to the differential equation in Eq.~\eqref{eq:summed_de} are obtained as:
\begin{align}
S_0(w)=& w^{\frac{2 n \gamma _0}{\beta _0}}\,,\\S_1(w)=& \left(T_{1,0}^X+\frac{2 n L_w \gamma _0 \left(\beta _1-2 \beta _0 \gamma _0\right)}{\beta _0^2}+\frac{2 n \left(\beta _1 \gamma _0-\beta _0 \gamma _1\right)}{\beta _0^2}\right) w^{\frac{2 n \gamma _0}{\beta _0}-1}+\frac{2 n \left(\beta _0 \gamma _1-\beta _1 \gamma _0\right) w^{\frac{2 n \gamma _0}{\beta _0}}}{\beta _0^2}\,,\\S_2(w)=&\frac{n w^{\frac{2 n \gamma _0}{\beta _0}}}{\beta _0^4} \Bigg[\gamma _2 \beta _0^3-\left(\beta _2 \gamma _0+\gamma _1 \left(\beta _1-2 n \gamma _1\right)\right) \beta _0^2+\beta _1 \gamma _0 \left(\beta _1-4 n \gamma _1\right) \beta _0+2 n \beta _1^2 \gamma _0^2\Bigg] \nonumber\\&+ w^{\frac{2 n \gamma _0}{\beta _0}-2}\Bigg[T_{2,0}^X+\frac{L_w \left(\beta _1-2 \beta _0 \gamma _0\right) \left(-\beta _0^2 \left(\beta _0-2 n \gamma _0\right) T_{1,0}^X+4 n \left(n \beta _1-\beta _0^2\right) \gamma _0^2+2 n \beta _0 \left(\beta _0-2 n \gamma _0\right) \gamma _1\right)}{\beta _0^4}\nonumber\\&\bs\bs+\frac{n}{\beta _0^4} \Big(\left(-2 \gamma _1 T_{1,0}^X+4 \gamma _0 \gamma _1-\gamma _2\right) \beta _0^3+\left(2 n \gamma _1^2-\beta _2 \gamma _0+\beta _1 \left(\gamma _1+2 \gamma _0 \left(T_{1,0}^X-2 \gamma _0\right)\right)\right) \beta _0^2\nonumber\\&\bs\bs+\beta _1 \gamma _0 \left(\beta _1-4 n \gamma _1\right) \beta _0+2 n \beta _1^2 \gamma _0^2\Big)-\frac{n L_w^2 \gamma _0 \left(\beta _0-2 n \gamma _0\right) \left(\beta _1-2 \beta _0 \gamma _0\right){}^2}{\beta _0^4}\Bigg] \nonumber\\&+w^{\frac{2 n \gamma _0}{\beta _0}-1}\Bigg[\frac{2 n}{\beta _0^4} \Big(\gamma _1 \left(T_{1,0}^X-2 \gamma _0\right) \beta _0^3+\left(\gamma _0 \left(\beta _2+\beta _1 \left(2 \gamma _0-T_{1,0}^X\right)\right)-2 n \gamma _1^2\right) \beta _0^2-\beta _1 \gamma _0 \left(\beta _1-4 n \gamma _1\right) \beta _0\nonumber\\&\bs\bs-2 n \beta _1^2 \gamma _0^2\Big)-\frac{4 n^2 L_w \gamma _0 \left(\beta _1-2 \beta _0 \gamma _0\right) \left(\beta _1 \gamma _0-\beta _0 \gamma _1\right)}{\beta _0^4}\Bigg]\,,\\
\end{align}
\begin{align}
S_3(w)= &\frac{2 n w^{\frac{2 n \gamma _0}{\beta _0}} }{3 \beta _0^6}S_{3,0}(w)+w^{\frac{2 n \gamma _0}{\beta _0}-1}S_{3,1}(w) +w^{\frac{2 n \gamma _0}{\beta _0}-2} S_{3,2}(w) +w^{\frac{2 n \gamma _0}{\beta _0}-3} S_{3,3} \,,
\end{align}
where,
\begin{align}
    S_{3,0}(w)&=\Bigg[\gamma _3 \beta _0^5-\left(\beta _3 \gamma _0+\beta _2 \gamma _1+\left(\beta _1-3 n \gamma _1\right) \gamma _2\right) \beta _0^4-\beta _1 \gamma _0 \left(\beta _1^2-6 n \gamma _1 \beta _1+6 n^2 \gamma _1^2-3 n \beta _2 \gamma _0\right) \beta _0^2\nonumber\\&+\left(\gamma _1 \beta _1^2+\left(2 \beta _2 \gamma _0-3 n \left(\gamma _1^2+\gamma _0 \gamma _2\right)\right) \beta _1+n \gamma _1 \left(2 n \gamma _1^2-3 \beta _2 \gamma _0\right)\right) \beta _0^3+3 n \beta _1^2 \gamma _0^2 \left(2 n \gamma _1-\beta _1\right) \beta _0-2 n^2 \beta _1^3 \gamma _0^3\Bigg]\,,
    \end{align}
    \begin{align}
    S_{3,1}(w)&=\Bigg[\frac{2 n^2 L_w \gamma _0 \left(\beta _1-2 \beta _0 \gamma _0\right)}{\beta _0^6} \left(\gamma _2 \beta _0^3-\left(\beta _2 \gamma _0+\gamma _1 \left(\beta _1-2 n \gamma _1\right)\right) \beta _0^2+\beta _1 \gamma _0 \left(\beta _1-4 n \gamma _1\right) \beta _0+2 n \beta _1^2 \gamma _0^2\right)\nonumber\\&\bs+\frac{n}{\beta _0^6} \bigg(\gamma _2 \left(T_{1,0}^X-2 \gamma _0\right) \beta _0^5+\beta _1 \gamma _0 \left(\beta _1^2+2 n \left(\gamma _0 T_{1,0}^X-4 \left(\gamma _0^2+\gamma _1\right)\right) \beta _1+6 n \left(2 n \gamma _1^2-\beta _2 \gamma _0\right)\right) \beta _0^2\nonumber\\&\bs+\left(-4 n^2 \gamma _1^3+6 n \beta _2 \gamma _0 \gamma _1+\beta _1^2 \gamma _0 \left(T_{1,0}^X-2 \gamma _0\right)+2 \beta _1 \left(n \left(8 \gamma _1 \gamma _0^2+\left(\gamma _2-2 \gamma _1 T_{1,0}^X\right) \gamma _0+\gamma _1^2\right)-\beta _2 \gamma _0\right)\right) \beta _0^3\nonumber\\&\bs+\left(-\left(\left(\beta _2 \gamma _0+\gamma _1 \left(\beta _1-2 n \gamma _1\right)\right) T_{1,0}^X\right)+\beta _3 \gamma _0+2 \gamma _0 \left(\beta _2 \gamma _0+\gamma _1 \left(\beta _1-4 n \gamma _1\right)\right)-2 n \gamma _1 \gamma _2\right) \beta _0^4\nonumber\\&\bs+6 n \beta _1^2 \gamma _0^2 \left(\beta _1-2 n \gamma _1\right) \beta _0+4 n^2 \beta _1^3 \gamma _0^3\bigg)\Bigg]\,,
    \end{align}
    \begin{align}
    S_{3,2}(w)&=\Bigg[-\frac{2 n^2 L_w^2 \gamma _0}{\beta _0^6}\Bigg( \left(\beta _0-2 n \gamma _0\right) \left(\beta _0 \gamma _1-\beta _1 \gamma _0\right) \left(\beta _1-2 \beta _0 \gamma _0\right)^2\Bigg)-\frac{2 n L_w \left(\beta _1-2 \beta _0 \gamma _0\right)}{\beta _0^6} \times\nonumber\\&\bs\bigg(\gamma _1 \left(T_{1,0}^X-2 \gamma _0\right) \beta _0^4+\left(\beta _2 \gamma _0+\beta _1 \left(2 \gamma _0-T_{1,0}^X\right) \gamma _0-2 n \gamma _1 \left(\gamma _1+\gamma _0 \left(T_{1,0}^X-4 \gamma _0\right)\right)\right) \beta _0^3+4 n^2 \beta _1^2 \gamma _0^3\nonumber\\&\bs-\gamma _0 \left(\beta _1^2+2 n \left(-\gamma _0 T_{1,0}^X+4 \gamma _0^2-\gamma _1\right) \beta _1+2 n \left(\beta _2 \gamma _0-2 n \gamma _1^2\right)\right) \beta _0^2+2 n \beta _1 \gamma _0^2 \left(\beta _1-4 n \gamma _1\right) \beta _0\bigg)\nonumber\\&\bs+\frac{1}{\beta _0^6}\bigg(2 \beta _0^6 \gamma _1 T_{1,0}^X-4 n^3 \beta _1^3 \gamma _0^3+6 n^2 \beta _0 \beta _1^2 \gamma _0^2 \left(2 n \gamma _1-\beta _1\right)+2 n^2 \beta _0^2 \beta _1 \gamma _0 \big(2 \beta _1 \left(-\gamma _0 T_{1,0}^X+4 \gamma _0^2+\gamma _1\right)\nonumber\\&\bs-6 n \gamma _1^2+3 \beta _2 \gamma _0\big)+2 n \beta _0^3 \big(\left(-\gamma _0 T_{1,0}^X+2 \gamma _0^2-\gamma _1\right) \beta _1^2+n \big(-16 \gamma _1 \gamma _0^2+\left(4 \gamma _1 T_{1,0}^X+\gamma _2\right) \gamma _0+\gamma _1^2\big) \beta _1\nonumber\\&\bs+n \gamma _1 \left(2 n \gamma _1^2-3 \beta _2 \gamma _0\right)\big)+\beta _0^5 \left(-\beta _2 T_{1,0}^X-2 \gamma _0 \left(\beta _1+2 n \gamma _1\right) T_{1,0}^X+8 n \gamma _0^2 \gamma _1+2 n \gamma _1 \left(T_{2,0}^X-2 \gamma _1\right)\right)\nonumber\\&\bs+\beta _0^4 \big(\beta _1^2 T_{1,0}^X+2 n^2 \gamma _1 \left(-2 \gamma _1 T_{1,0}^X+8 \gamma _0 \gamma _1-\gamma _2\right)-2 n \beta _1 \gamma _0 \left(-2 \gamma _0 T_{1,0}^X+T_{2,0}^X+4 \gamma _0^2-2 \gamma _1\right)\nonumber\\&\bs+2 n \beta _2 \left(\gamma _1+\gamma _0 \left(T_{1,0}^X-2 \gamma _0\right)\right)\big)\bigg)\Bigg] \,,
    \end{align}
    \begin{align}
     S_{3,3}(w)&=\Bigg[\frac{2 n L_w^3 \gamma _0 \left(\beta _0-n \gamma _0\right) \left(\beta _0-2 n \gamma _0\right) \left(\beta _1-2 \beta _0 \gamma _0\right)^3}{3 \beta _0^6}-\frac{L_w \left(\beta _1-2 \beta _0 \gamma _0\right) }{\beta _0^6}\Big(2 \left(T_{2,0}^X-\gamma _0 T_{1,0}^X\right) \beta _0^5\nonumber\\&\bs+\left(\beta _1 T_{1,0}^X-2 n \left(-2 \gamma _0^2 T_{1,0}^X+2 \gamma _1 T_{1,0}^X+4 \gamma _0^3+\gamma _2+\gamma _0 \left(T_{2,0}^X-6 \gamma _1\right)\right)\right) \beta _0^4+2 n \big(\gamma _0 \left(\beta _1+2 n \gamma _1\right) T_{1,0}^X\nonumber\\&\bs-2 \beta _1 \gamma _0^2+2 n \gamma _1^2-\beta _2 \gamma _0-8 n \gamma _0^2 \gamma _1+n \gamma _0 \gamma _2\big) \beta _0^3+2 n^2 \beta _1 \gamma _0^2 \left(4 n \gamma _1-\beta _1\right) \beta _0\nonumber\\&\bs+2 n \gamma _0 \left(\beta _1^2+n \left(-2 \gamma _0 T_{1,0}^X+8 \gamma _0^2-3 \gamma _1\right) \beta _1+n \left(\beta _2 \gamma _0-2 n \gamma _1^2\right)\right) \beta _0^2-4 n^3 \beta _1^2 \gamma _0^3\Big)\nonumber\\&\bs+\frac{L_w^2 \left(\beta _1-2 \beta _0 \gamma _0\right)^2}{\beta _0^6} \Big(\beta _0^2 \left(\beta _0-2 n \gamma _0\right) \left(\beta _0-n \gamma _0\right) T_{1,0}^X-n \big(2 \left(\gamma _1-3 \gamma _0^2\right) \beta _0^3+\gamma _0 \left(8 n \gamma _0^2+\beta _1-6 n \gamma _1\right) \beta _0^2\nonumber\\&\bs+2 n \gamma _0^2 \left(\beta _1+2 n \gamma _1\right) \beta _0-4 n^2 \beta _1 \gamma _0^3\big)\Big)+\frac{1}{3 \beta _0^6}\big[3 \left(T_{3,0}^X-2 \gamma _1 T_{1,0}^X\right) \beta _0^6+\big(3 \left(\beta _2-n \gamma _2\right) T_{1,0}^X-6 n \gamma _1 T_{2,0}^X\nonumber\\&\bs+12 n \gamma _1^2-24 n \gamma _0^2 \gamma _1-2 n \gamma _3+6 \gamma _0 \left(\left(\beta _1+2 n \gamma _1\right) T_{1,0}^X+n \gamma _2\right)\big) \beta _0^5+\beta _0^4\big(-3 \beta _1^2 T_{1,0}^X-12 n \beta _1 \gamma _0^2 T_{1,0}^X\nonumber\\&\bs+6 n^2 \gamma _1^2 T_{1,0}^X+3 n \beta _1 \gamma _1 T_{1,0}^X+6 n \beta _1 \gamma _0 T_{2,0}^X+24 n \beta _1 \gamma _0^3-24 n^2 \gamma _0 \gamma _1^2-n \beta _3 \gamma _0-18 n \beta _1 \gamma _0 \gamma _1+2 n \beta _1 \gamma _2\nonumber\\&\bs+6 n^2 \gamma _1 \gamma _2+n \beta _2 \left(-3 \gamma _0 T_{1,0}^X+6 \gamma _0^2-4 \gamma _1\right)\big) +n \beta _0^3\big(-4 n^2 \gamma _1^3+6 n \beta _2 \gamma _0 \gamma _1+\beta _1^2 \left(4 \gamma _1+3 \gamma _0 \left(T_{1,0}^X-2 \gamma _0\right)\right)\nonumber\\&\bs+2 \beta _1 \left(\beta _2 \gamma _0+3 n \left(8 \gamma _1 \gamma _0^2-\left(2 \gamma _1 T_{1,0}^X+\gamma _2\right) \gamma _0-\gamma _1^2\right)\right)\big) +n \beta _1 \gamma _0\beta _0^2 \big(-\beta _1^2+6 n \gamma _0 \left(T_{1,0}^X-4 \gamma _0\right) \beta _1\nonumber\\&\bs+6 n \left(2 n \gamma _1^2-\beta _2 \gamma _0\right)\big) +6 n^2 \beta _1^2 \gamma _0^2 \left(\beta _1-2 n \gamma _1\right) \beta _0+4 n^3 \beta _1^3 \gamma _0^3\big]\Bigg]
\end{align}
where we have taken the $T_{0,0}^X=1$ in Eq.~\eqref{eq:mom_fopt}. 
\end{widetext}

\end{document}